\documentclass[12pt,preprint]{aastex}

\usepackage{amsmath}
\usepackage{amssymb}

\shorttitle{Transit Model Fitting and Multiple-Planet Search}
\shortauthors{J. Li et al.}

\makeatletter
\renewcommand{\@make@caption@text}[2]{%
  \begin{center}
    \makebox[\textwidth]{\rmfamily#1.\quad#2}
  \end{center}
}%
\makeatother

\begin{document}

% ************************************************************************

\title{\textit{Kepler} Data Validation {II} -- Transit Model Fitting and Multiple-Planet Search}

\author{Jie Li\altaffilmark{*,1}, Peter Tenenbaum\altaffilmark{1}, Joseph D. Twicken\altaffilmark{1}, Christopher J. Burke\altaffilmark{1}, Jon M. Jenkins\altaffilmark{2}, Elisa V. Quintana\altaffilmark{3}, Jason F. Rowe\altaffilmark{1}, Shawn E. Seader\altaffilmark{4}}

\altaffiltext{*}{email: jie.li-1@nasa.gov}
\altaffiltext{1}{SETI Institute, Mountain View, CA, 94043}
\altaffiltext{2}{NASA Ames Research Center, Moffett Field, CA, 94035}
\altaffiltext{3}{NASA Goddard Space Flight Center, Greenbelt, MD, 20771}
\altaffiltext{4}{Rincon Research Corporation, Tucson, AZ, 85711}


% ************************************************************************

\begin{abstract}
This paper discusses the transit model fitting and multiple-planet search algorithms and performance of the \textit{Kepler} Science Data Processing Pipeline, developed by the \textit{Kepler} Science Operations Center (SOC). Threshold Crossing Events (TCEs), which are transit candidate events, are generated by the Transiting Planet Search (TPS) component of the pipeline and subsequently processed in the Data Validation (DV) component. The transit model is used in DV to fit TCEs in order to characterize planetary candidates and to derive parameters that are used in various diagnostic tests to classify them. After the signature associated with the TCE is removed from the light curve of the target star, the residual light curve goes through TPS again to search for additional TCEs. The iterative process of transit model fitting and multiple-planet search continues until no TCE is generated from the residual light curve or an upper limit is reached. The transit model fitting and multiple-planet search performance of the final release (9.3, January 2016) of the pipeline is demonstrated with the results of the processing of 4 years (17 quarters) of flight data from the primary \textit{Kepler Mission}. The transit model fitting results are accessible from the NASA Exoplanet Archive. The final version of the SOC codebase is available through GitHub.
\end{abstract}

% ***********************************************************************************************************
%
%						Introduction
%
% ***********************************************************************************************************
\section{Introduction}

This paper discusses transit model fitting and multiple-planet search algorithms and performance that are part of the Data Validation (DV) component of the \textit{Kepler} Science Data Processing Pipeline \citep{jenkins2010a}, developed by the \textit{Kepler} Science Operations Center (SOC) at NASA Ames Research Center. An introduction to \textit{Kepler Mission}, the \textit{Kepler} Science Data Processing Pipeline and the DV component is provided in a companion paper \citep{jdt2018}, which also details the DV diagnostic tests and data products for vetting transiting planet candidates. 

The transit model fitting is designed for the following three main tasks: (1) The orbital property and the nature of the planetary candidates are characterized; (2) The fitted parameters of the transit model and the corresponding light curve generated from the model are used in the diagnostic tests in DV to aid in the assessment and classification of planetary candidates; (3) When the Transiting Planet Search (TPS) component is called, only one Threshold Crossing Event (TCE) with the maximum multiple event detection statistic (MES) is generated. To search for multiple-planetary candidates, an iterative process of transit model fitting and multiple-planet search is implemented in DV. For each target star, the transit model parameters are fitted to each TCE generated by TPS, the signature of known TCEs is removed from the light curve, and then the residual is provided to TPS again to search for additional TCEs. This iteration will only terminate once no new TCEs are identified or a preset upper limit is reached (set to 10 for the SOC 9.3 run producing the Data Release (DR) 25 TCEs).

The transit model fitting results, such as the fitted parameters and uncertainties, derived parameters and uncertainties, goodness of fit metrics, and the diagnostic plots, are included in comprehensive DV reports by target, and one-page DV summary reports by TCE. The reports and summaries are accessible by the science community at the Exoplanet Archive\footnote{https://exoplanetarchive.ipac.caltech.edu.} at the NASA Exoplanet Science Institute (NExScI) \citep{akeson2013}. The final version of the SOC 9.3 codebase is available to the general public through GitHub\footnote{https://github.com/nasa/kepler-pipeline.}.

The transit model fitting and multiple-planet search algorithm in the initial revision of DV (SOC 6.1) was described by \citet{pt2010} and \citet{wu2010}. DV evolved greatly since then. Major changes in the transit model fitting and multiple-planet search algorithm include: (1) The transit model described in \citet{pt2010} and \citet{wu2010} is changed to the geometric transit model, including a nonlinear limb-darkening model \citep{claret2011}, for a better modeling accuracy; (2) The reduced parameter fits are added; (3) The trapezoidal model fit is added. 

Iterative transit fitting and multiple-planet search has been done extensively by various groups. \citet{foreman2015} performs a joint fit of the transit model and systematics, which may be more sensitive than the algorithm used in the SOC 9.3 codebase but is computationally more expensive. \citet{crossfield2016}, \citet{crossfield2018}, \citet{petigura2018}, and \citet{yu2018} use the ``TERRA" software package and presume the systematic error correction has whitened the colored noise of the light curve. \citet{dressing2015}, \citet{vanderburg2016}, and \citet{rizzuto2017} use box least squares algorithm and also assume that the residual observation noise is white. In this paper, the transit model fitting is implemented with an iterative loop that includes a whitening filter and a transit fitter. In addition, compared to the similar work by other groups,  the reduced parameter fits described in this paper improve the consistency of the results of the geometric transit model fit, and the trapezoidal model fit provides a quick assessment of the transit signal. 

In this paper, the final SOC 9.3 codebase is described. The architecture of transit model fitting and multiple-planet search algorithm is described in Section \ref{sec:Architecture}, and the light curve preprocessing procedures are described in Section \ref{sec:LightCurvePreprocessing}. The geometric transit model is described in Sections \ref{sec:GeometricTransitModel}. Section \ref{sec:GeometricTransitSignalGenerator} describes how a synthetic light curve is generated from the fitted parameters of the geometric transit model, and Section \ref{sec:GeometricModelFitting} describes the algorithms to fit the light curves with the geometric transit model. A fitting algorithm with the trapezoidal model is described in Section \ref{sec:TrapezoidalModelFitting}, and the multiple-planet search is discussed in Section  \ref{sec:MultiplePlanetSearch}.  The performance of the transit model fitting and multiple-planet search is demonstrated in Section \ref{sec:PerformanceOfTransitModelFittingAndMultiplePlanetSearch}. Finally, conclusions are presented in Section \ref{sec:Conclusions}.

% ***********************************************************************************************************
%
%						Architecture of Transit Model Fitting and Multiple-Planet Search
%
% ***********************************************************************************************************
\section{Architecture of Transit Model Fitting and Multiple-Planet Search} 
\label{sec:Architecture}

\begin{figure}
	\centering{
		\includegraphics[width=0.9\textwidth]{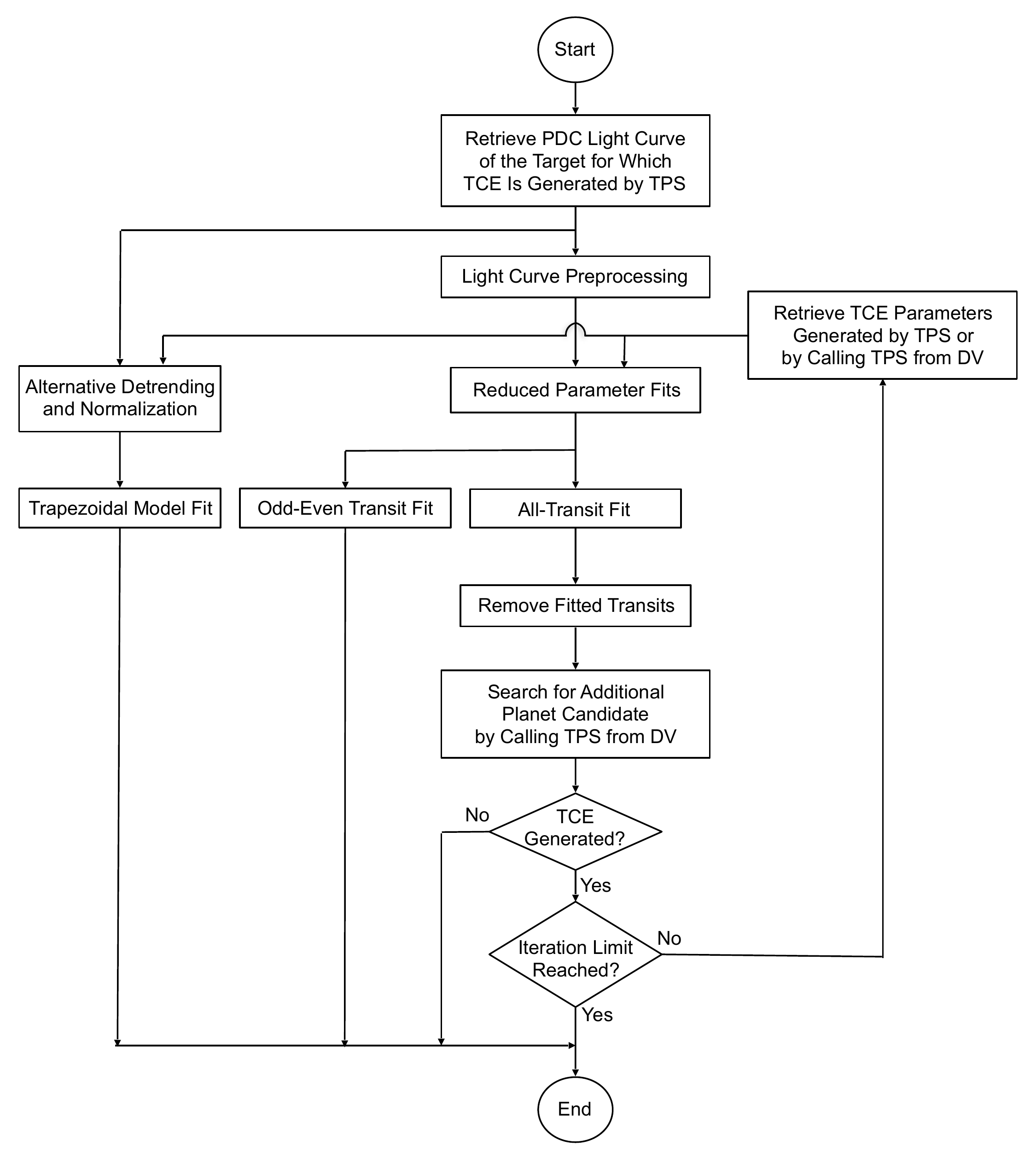}
		\caption{Flowchart of iterative process of transit model fitting and multiple-planet search. In the flowchart, a rectangle represents an operation of data processing, a diamond  represents a conditional operation that determines which one of the two paths the process will take, an arrow line shows the order of operations, and an oval represents the beginning or ending of the process.}
		\label{fig:flowchart-transit-model-fitting-and-multiple-planet-search}}
\end{figure}

This section describes the architecture of transit model fitting and multiple-planet search algorithm. As shown in the flowchart in 
Figure \ref{fig:flowchart-transit-model-fitting-and-multiple-planet-search}, it is an iterative process. 

When a TCE is generated by the TPS component, the corresponding systematic error-corrected light curve of the target star, generated by the Presearch Data Conditioning (PDC) component of the pipeline, is furnished to DV along with the transit parameters associated with the TCE, including the transit epoch (central time of first transit), orbital period, transit duration, and MES of the TCE. The light curve may span one or more observing quarters. After several preprocessing procedures, the light curve of the target star goes through a series of transit model fitting algorithms, which include reduced parameter fits, all-transit fit, odd-even transit fit and trapezoidal model fit. 

As shown in Figure \ref{fig:flowchart-transit-model-fitting-and-multiple-planet-search}, the preprocessed light curve is first subjected to a set of reduced parameter fits, in which the impact parameter is set to fixed values of 0.1, 0.3, 0.5, 0.7 and 0.9, and only the parameters of transit epoch time, planet orbital period, ratio of planet radius to star radius and ratio of semi-major axis to star radius are fitted to a geometric transit model. The initial values of the fitted parameters of the reduced parameter fits are determined from the TCE parameters. The reduced parameter fits resolve the degenerate problem of fitting the impact parameter, which is discussed in Subsection \ref{ssec:reducedParameterFits}. After the completion of the reduced parameter fits, all-transit fit and odd-even transit fit follow, in which the fitting algorithms are applied to all transits, odd transits and even transits, respectively. The all-transit fit and odd-even transit fit are both initialized with the fitted parameters of the reduced parameter fit with the minimum $\chi^2$ metric. The output of the all transit fit is used in several diagnostic tests of DV and the assessment of planet candidacy, and the output of the odd-even transit fit is used in a specific DV diagnostic test to identify false positives due to an eclipsing binary target or a target with an eclipsing binary in the background. In addition to the fitting algorithms with the geometric transit model, a fitting algorithm with the trapezoidal model is implemented. As shown in Figure \ref{fig:flowchart-transit-model-fitting-and-multiple-planet-search}, an alternative detrending and normalization algorithm is applied to the PDC light curve prior to the trapezoidal model fit. The output of the trapezoidal model fit is used in the diagnostic tests of DV when the fit with the geometric transit model fails or when the fit is not performed, e.g. for suspected eclipsing binaries based on transit depth.

After the completion of the transit model fitting algorithms, the signature of the known TCE, as determined from the fitted parameters of the all-transit fit, is removed from the light curve, and the residual light curve is subjected to a search for additional planets by calling TPS in the DV component. If an additional TCE is generated, the residual light curve goes through the transit model fitting algorithms discussed above once again. The iterative process of the transit model fitting and multiple-planet search concludes when no additional TCEs are produced or an upper limit is reached.

% ***********************************************************************************************************
%
%						Light Curve Preprocessing
%
% ***********************************************************************************************************
\section{Light Curve Preprocessing}
\label{sec:LightCurvePreprocessing}

The light curves of target stars are processed in the PDC component before they are input in the DV component. As described in \citet{stumpe2012} and \citet{js2012}, systematic errors due to the thermal transients and optical distortions are estimated and compensated, outliers due to cosmic rays and transients due to the Argabrightening events\footnote{An Argabrightening event, which was described by \citet{witteborn2011}, is an occasional diffuse illumination of portions of the focal plane lasting a few minutes.} are removed, and sudden pixel sensitivity dropouts are identified and corrected. Nevertheless, the PDC light curves must be preprocessed further in preparation for transit model fitting. The preprocessing procedures in DV include baseline removal, light curve normalization, quarterly data segment stitching, harmonic removal, and timestamp conversion.

\subsection{Baseline Removal and Light Curve Normalization}
\label{ssec:baselineRemovalAndNormalization}

The light curve generated in the PDC component measures the brightness of the target star in units of photo-electrons (e$^-$) per cadence\footnote{The flux units in the \textit{Kepler} light curve files exported to the Mikulski Archive for Space Telescopes (MAST) are actually e$^-$/second. The conversion from e$^-$/cadence to e$^-$/second is performed in the Archive (AR) component of the pipeline.}. Since the brightness of one target star is generally measured by four different charge coupled device (CCD) channels over the course of a year due to the quarterly rotations of the spacecraft about the telescope boresight, the baseline of the measured light curve of the target star varies from quarter to quarter. The preprocessing procedure of baseline removal and light curve normalization removes the baseline of the measurement and generates the normalized light curves so that they can later be uniformly processed by the transit model fitting algorithms. This preprocessing procedure is implemented quarter by quarter in two steps: (1) For each target star, the median flux level is determined for each quarter.  For stars on the same CCD channel, the median flux level varies from one target star to another depending on the magnitude and spectrum of the target star; for a given target star, the median varies from quarter to quarter depending on the sensitivity of the CCD pixels and the sub-pixel location of the stellar image.  (2) The median is subtracted from the corresponding quarterly light curves and the difference is normalized by the median and multiplied by 10$^6$, yielding a normalized light curve in units of parts per million (ppm). For the out-of-transit data points, the baseline value is zero. For in-transit data points, the normalized flux is negative and its absolute value corresponds to the ratio of the flux blocked by the transiting planet to the total flux of the target star. For example, to an extraterrestrial observer of a central transit, the depth of the normalized light curve of the Earth transiting the Sun is about $84\times10^{-6}$, or 84 ppm.

\subsection{Quarterly Data Segment Stitching}
\label{ssec:dataSegmentStitch}

The light curve of a target star is comprised of data segments separated by gaps that may have resulted from quarterly rolls, monthly data downlinks, or spacecraft anomalies. The preprocessing procedure of data segment stitching removes the trend and transients of the light curve of the segment edges and fills the gaps between the segments. The trend of the light curve of each segment is identified, and the light curve at the edges of the segments, where transients are usually observed, is fitted with a model of linear and exponential components. Then the detrending algorithm removes the identified trend and the fitted components.  The gaps between the data segments are filled with different methods, depending on the length of the gaps: the short gaps are filled with an auto-regressive model and the long gaps are filled via data reflection and tapering \citep{jenkins2002, jenkins2010b, jenkins2017}.

\subsection{Harmonic Removal}
\label{ssec:harmonicRemoval}

The harmonic removal procedure identifies and removes sinusoidal harmonic components, which are significant in the light curves of target stars such as rotating and contact binaries. The light curve is first processed with a Fast Fourier Transform (FFT) to determine the frequencies of the significant harmonic components. Then the magnitude and phase of the components are fitted and the significant harmonic components are removed from the light curve \citep{jenkins2002, jenkins2010b, jenkins2017}. It is possible that the harmonic removal process may degrade the transits of short-period planets. This was discussed by \citet{christiansen2013, christiansen2015}.

\subsection{Timestamp Conversion}
\label{ssec:timestampConversion}

Based on Kepler's laws of planetary motion, the transits of exoplanets are inherently periodic\footnote{This neglects transit timing variations, which can be quite large for dynamically packed planetary systems with planets in near-orbital resonances.} if the observer is located at the barycenter of the Solar System and the events are measured in the Barycentric Dynamical Time (TDB) frame. However, the timestamps associated with the PDC light curves provided to DV are Modified Julian Day (MJD) numbers, which correspond to the time when the light of the target star arrives at the \textit{Kepler} spacecraft in the Coordinated Universal Time (UTC) frame. Before the transit model fitting algorithms are applied, barycentric time corrections are applied to obtain timestamps in Barycentric Modified Julian Day (BMJD) numbers, to correspond to the time when the light from the events of the target star would arrive at the barycenter of the Solar System in the TDB frame. 

The algorithm to determine each BMJD timestamp requires the following inputs: the ephemeris of the \textit{Kepler} spacecraft, the ephemeris of the Solar System, and the celestial coordinates of the target star. Then the difference between the time when the light of the events of the target star would have reached the barycenter of the Solar System and the time when the same light arrived at the \textit{Kepler} spacecraft, which is located in an Earth-trailing heliocentric orbit, is calculated. Finally, the BMJD timestamps are determined as the sum of the MJD timestamps and the aforementioned barycentric time corrections. To simplify the processing and storage of the \textit{Kepler} science data, a new timestamp, Barycentric \textit{Kepler} Julian Date (BKJD), is defined and used in the \textit{Kepler} Science Data Processing Pipeline and the NASA Exoplanet Archive. By definition, BKJD is equal to BMJD minus a constant of 54,832.5 days, which corresponds to 12:00:00 noon on January 1, 2009 (the first day of the year when the \textit{Kepler} spacecraft was launched). After the preprocessing procedure of timestamp conversion, all light curves are associated with BKJD timestamps. The time frames and the timestamps before and after timestamp conversion in the preprocessing are summarized in Table \ref{table:timestamp-conversion}.

\begin{table}
	\centering{
		\caption{Time frames and timestamps before and after the timestamp conversion.}
		\begin{tabular}{ccc}
			\textbf{Before/After Conversion} & \textbf{Time Frame} & \textbf{Timestamp}    \\
			\hline
			before                          & UTC                 & MJD 	              \\
			after                           & TDB                 & BKJD (=BMJD-54,832.5) \\
		\end{tabular}
		\label{table:timestamp-conversion}}
\end{table}

\begin{figure}
	\centering{
		\includegraphics[width=0.8\textwidth]{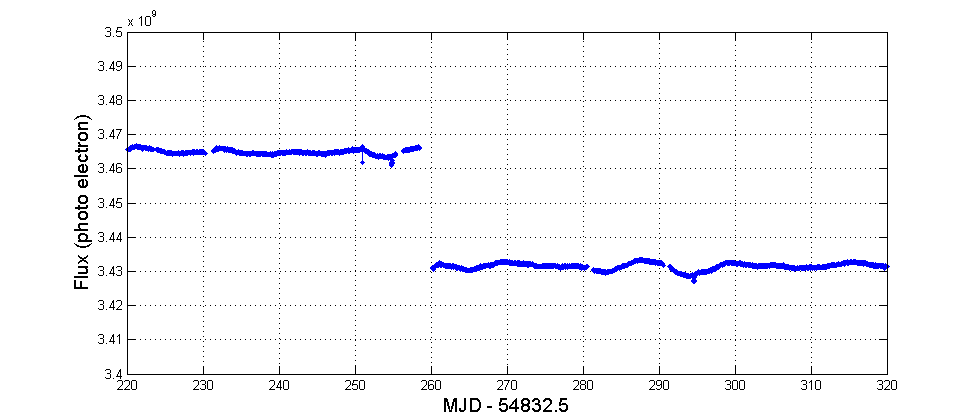}
		\includegraphics[width=0.8\textwidth]{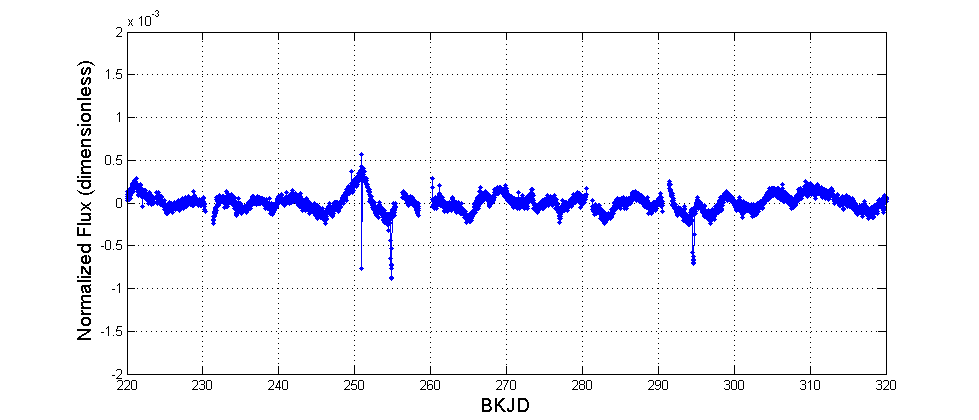}
		\caption{Flux time series of KIC 8478994 before (top) and after (bottom) the preprocessing procedures.}
		\label{fig:light-curve-preprocessing-flux}}
\end{figure}

As an example, Figure \ref{fig:light-curve-preprocessing-flux} shows two segments of the light curve of the target star KIC 8478994,  or Kepler-37, before and after the preprocessing procedures. As illustrated in the figure, the light curve before the preprocessing shows the absolute flux value in units of photo-electrons, timestamped in MJD, and the light curve after the preprocessing shows the dimensionless normalized flux value, timestamped in BKJD.

% ***********************************************************************************************************
%
%						Geometric Transit Model
%
% ***********************************************************************************************************
\section{Geometric Transit Model}
\label{sec:GeometricTransitModel}

The transit model fitting algorithms of the DV component employ the geometric transit model of \citet{mandel2002}, including a nonlinear limb-darkening model, parameterized as per \citet{claret2011}.  The limb-darkening depends on the stellar parameters, such as radius $R_s$ (solar radii), surface gravity $\log{g}$ ($\log_{10}$(cm s$^{-2}$)), metallicity $\log_{10}$[M/H] (dimensionless), and effective temperature $T_{eff}$ (K), which are extracted from the Kepler Input Catalog (KIC) \citep{brown2011} or override to KIC parameter values \citep{mathur2017}.

\subsection{Fitted Parameters}
\label{ssec:fittedParameters}

In the geometric transit model, the eccentricity and the longitude of periapsis of the planet orbit around the host star are assumed to be 0, and the five parameters to be fitted are listed below:

\begin{itemize}
	\item Transit epoch time $t_{epoch}$ (BKJD): the time corresponding to the center of the first detected transit;
	\item Orbital period $P$ (days): the interval between consecutive planetary transits, i.e., the period of the planet's orbit;
	\item Ratio of planet radius to stellar radius $R_{p}/R_{s}$ (dimensionless): the ratio of the planet radius divided by the stellar radius;
	\item  Ratio of semi-major axis to stellar radius $a/R_{s}$ (dimensionless): since the eccentricity of the planet's orbit is assumed to be zero, this is the ratio of the distance between the planet and the host star divided by the stellar radius;
	\item Impact parameter $b$ (dimensionless): the sky-projected distance between the center of the stellar disc and the center of the planet disc at conjunction, normalized by the stellar radius.
\end{itemize}

As discussed in Subsection \ref{sssec:LMFit}, the fitted parameters are determined with the iterative Levenberg-Marquardt (LM) algorithm \citep{levenberg1944, marquardt1963}. The all-transit fit and odd-even transit fit are both initialized with the fitted parameters of the reduced parameter fit with the minimum $\chi^2$ metric. 

In the reduced parameter fits, the impact parameter $b$ is set to fixed values of 0.1, 0.3, 0.5, 0.7 and 0.9. The initial values of the fitted parameters $t_{epoch}$, $P$, $R_{p}/R_{s}$ and $a/R_{s}$ are determined from the TCE parameters provided by the TPS component. The TPS value for orbital period can be used directly. Note that the transit epoch time from the TPS component is in units of MJD, while the fitted parameter of $t_{epoch}$ is in units of BKJD; therefore, a unit conversion is required. The initial values of $R_{p}/R_{s}$ and  $a/R_{s}$  are determined from the TCE parameters according to: 

\begin{equation}
\frac{R_p}{R_s} = \left(\frac{SES_{TPS}}{r_{flux}}\right)^{\frac{1}{2}} \left( \frac{d_{lc}}{d_{tr,TPS}} \right)^{\frac{1}{4} }\; \mathrm{and}
\end{equation}

\begin{equation}
\frac{a}{R_s} = \left( \frac{\left(1+\frac{R_p}{R_s}\right)^2 - b^2}{\sin^2\left(\frac{\pi \, d_{tr,TPS}}{24 \, P_{TPS}}\right)} +b^2 \right)^{\frac{1}{2}},
\end{equation}

\noindent where the single event statistic $SES_{TPS}$ (dimensionless),  orbital period $P_{TPS}$ (days), and transit duration $d_{tr,TPS}$ (hours) are TCE parameters determined in the TPS component. $r_{flux}$ is the ratio of the light curve value to the uncertainty. $d_{lc}$ (hours) is the duration of a long-cadence (LC) interval (29.4 min or 0.49 hr).

\subsection{Derived Parameters}
\label{ssec:derivedParameters}

Once the transit model fitting algorithm has converged, several additional parameters regarding the planet or the transits can be derived from the fitted parameters. 

Given the stellar radius $R_s$, the planet radius $R_p$ is determined directly from the fitted parameter $R_p/R_s$:

\begin{equation}\label{equation:planet-radius}
R_p = \left(\frac{R_{\odot}}{R_{\oplus}}\right) \left(\frac{R_p}{R_s}\right) R_s,
\end{equation}

\noindent where $R_{\odot}$ and $R_{\oplus}$ are radii of the Sun and the Earth, respectively, both in units of $m$. Since $R_s$ is in units of solar radii, $R_p$ given by Equation \ref{equation:planet-radius} is in units of Earth radii.

Before calculating the semi-major axis of the planet orbit $a$, the planet-star separation for a circular orbit, the acceleration due to gravity at the surface of the star $g$ should first be determined from the stellar parameter $\log{g}$ as: 

\begin{equation}\label{equation:surface-gravity}
g=\frac{1}{100} \, 10^{\log{g} }.
\end{equation}

\noindent A factor of 1/100 is required to convert acceleration $g$ to units of m s$^{-2}$ from $\log{g}$ in units of $\log_{10}$(cm s$^{-2}$).

The semi-major axis of the planetary orbit $a$ is not determined directly from the fitted parameter $a/R_s$, but derived from the fitted orbital period $P$ based on Kepler's third law:

\begin{equation}\label{equation:semi-major-axis}
%a = \left(\frac{1}{f_{AU}}\right) \left(\frac{\left(86400 \, P\right) \, \left(R_s \, R_{\odot}\right)  \sqrt{g}}{2 \pi}\right)^\frac{2}{3},
a = \frac{1}{f_{AU}} \left[\frac{\left(86400 \, P\right) \, \left(R_s \, R_{\odot}\right)  \sqrt{g}}{2 \pi}\right]^\frac{2}{3},
\end{equation}

\noindent where $f_{AU}$ is the factor to convert the astronomical unit (AU) to m (i.e., the number of m in one AU). Please note $P$ and $R_s$ are in units of days and solar radii, respectively, so Equation \ref{equation:semi-major-axis} gives the semi-major axis of the planet orbit, $a$, in AU. 

The inclination of the planet orbit $i$ in units of degrees, is determined from fitted parameters $b$ and $a/R_s$:

\begin{equation}
i = \frac{180}{\pi} \, \cos^{-1}\left(\frac{b}{a/R_s}\right).
\end{equation}

\begin{figure}
	\centering{
		\includegraphics[width=15 cm]{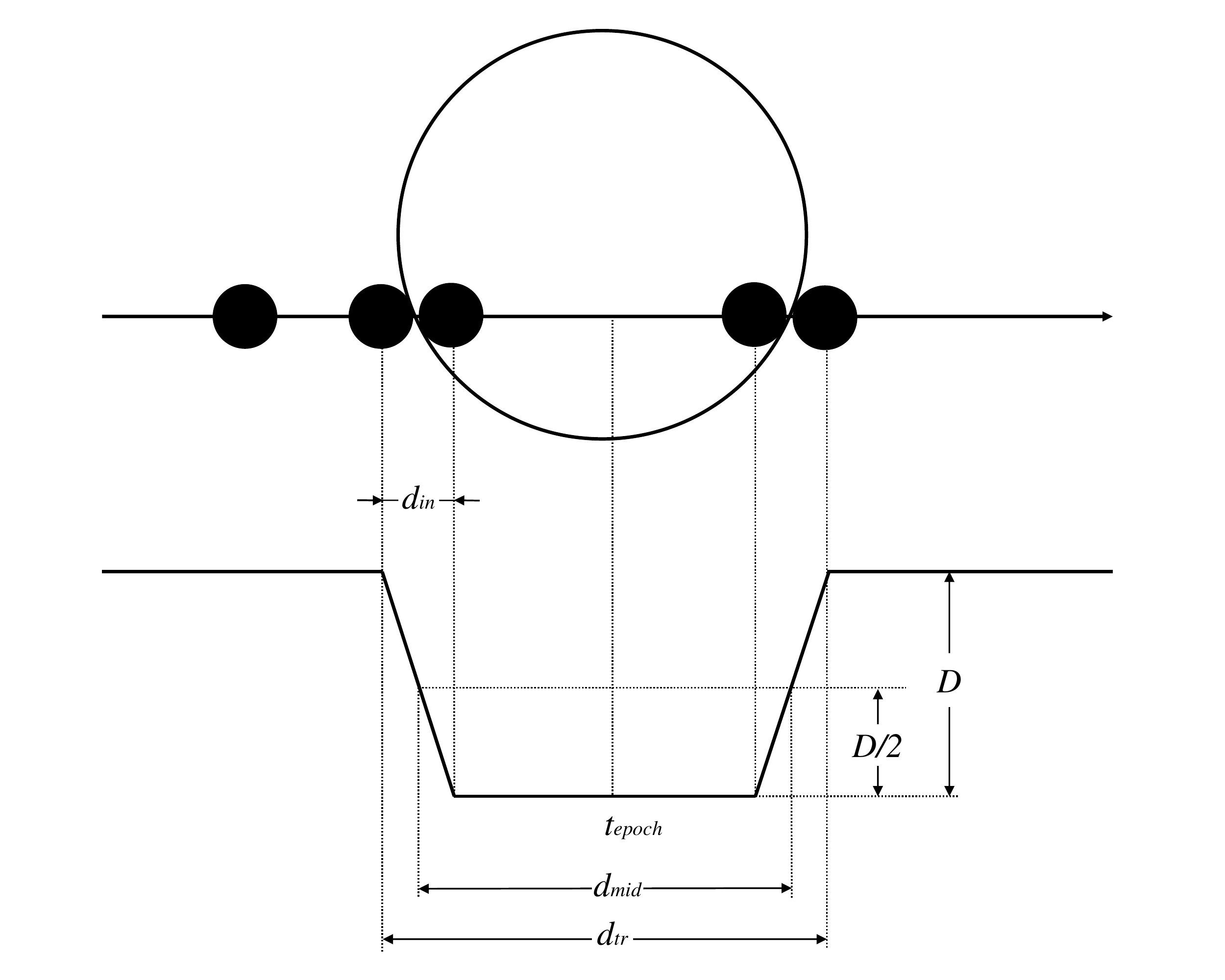}
		\caption{Schematic of planetary transit and associated light curve with the depth, duration, ingress time, and epoch time indicated.}
		\label{fig:transit-model-diagram}}
\end{figure}

As illustrated in Figure \ref{fig:transit-model-diagram}, the transit depth $D$, transit duration $d_{tr}$, and transit ingress time $d_{in}$ are another set of parameters defining the size and shape of a transit. The transit depth $D$ is determined as the absolute value of the minimum of the normalized light curve generated from the fitted parameters (to be discussed in Section \ref{sec:GeometricTransitSignalGenerator}), multiplied by a factor of $10^6$ to convert the dimensionless normalized flux value to the transit depth in units of ppm. The parameters $d_{tr}$ and $d_{in}$, both in units of hours, are derived from fitted parameters $R_p/R_s$, $a/R_s$, $b$, and $P$ with the following equations:

\begin{equation}\label{transit-duration}
d_{tr} = \frac{24 \, P}{\pi} \, \sin^{-1} \left(\sqrt{\frac{\left(1+\frac{R_p}{R_s}\right)^2 - b^2}{\left(\frac{a}{R_s}\right)^2 - b^2}}\right), \, \mathrm{and}
\end{equation}

\begin{equation}
d_{in} = \frac{12 \, P}{\pi} \, \sin^{-1} \left(\sqrt{\frac{\left(1+\frac{R_p}{R_s}\right)^2 - b^2}{\left(\frac{a}{R_s}\right)^2 - b^2}} - \sqrt{\frac{\left(1-\frac{R_p}{R_s}\right)^2 - b^2}{\left(\frac{a}{R_s}\right)^2 - b^2}}\right).
\end{equation}

The planet equilibrium temperature $T_{eq}$, an estimate of the surface temperature of the planet, is calculated assuming a thermodynamic equilibrium is reached between the incident stellar flux and the radiated heat from the planet:

\begin{equation}
T_{eq} = T_{eff} \, \left(1-\alpha\right)^{\frac{1}{4}} \, \sqrt{\frac{R_s \, R_{\odot}}{2 \, a \, f_{AU}}} ,
\end{equation}

\noindent where $a$ is the semi-major axis of the planetary orbit in AU determined by Equation \ref{equation:semi-major-axis}, $\alpha$ is the albedo of the planet, whose default value is set to 0.3, and both $T_{eff}$ and $T_{eq}$ are in K.

The planet effective stellar flux, $\phi_{eff}$, defined as the ratio of the flux of the host star at the top of planet's atmosphere to the solar flux at the top of Earth's atmosphere, is calculated as

\begin{equation}
\phi_{eff} = \left(\frac{R_s}{a}\right)^2 \, \left(\frac{T_{eff}}{T_{eff, \, \odot}}\right)^4,
\end{equation}

\noindent where $a$ is determined by Equation \ref{equation:semi-major-axis} and $T_{eff, \, \odot}$ is the effective temperature of the Sun in units of K.

The fitted and derived parameters of the transit model are summarized in Table \ref{table:transit-model-parameters}.

\begin{table}
	\centering{
		\caption{Fitted and derived parameters of the transit model.}
		\begin{tabular}{cccc}
			\textbf{Parameter}                      & \textbf{Symbol} & \textbf{Unit} & \textbf{Fitted/Derived}  \\
			\hline
			transit epoch time                      & $t_{epoch}$     & BKJD          & fitted                   \\
			planet orbital period                   & $P$             & day           & fitted                   \\
			ratio of planet radius to star radius   & $R_p/R_s$       & dimensionless & fitted                   \\
			ratio of semi-major axis to star radius & $a/R_s$         & dimensionless & fitted                   \\     
			impact parameter                        & $b$             & dimensionless & fitted                   \\
			planet radius                           & $R_p$           & Earth radius  & derived                  \\
			planet orbit semi-major axis            & $a$             & AU            & derived                  \\
			planet orbit inclination                & $i$             & degree        & derived                  \\
			transit depth                           & $D$             & ppm           & derived                  \\
			transit duration                        & $d_{tr}$        & hour          & derived                  \\
			transit ingress time                    & $d_{in}$        & hour          & derived                  \\
			planet equilibrium temperature          & $T_{eq}$        & K           & derived                  \\
			planet effective stellar flux           & $\phi_{eff}$    & dimensionless & derived                  \\
		\end{tabular}
		\label{table:transit-model-parameters}}
\end{table}

% ***********************************************************************************************************
%
%						Geometric Transit Signal Generator
%
% ***********************************************************************************************************
\section{Geometric Transit Signal Generator}
\label{sec:GeometricTransitSignalGenerator}

The geometric transit signal generator generates a light curve at an array of cadence timestamps in BKJD (nominally the timestamps corresponding to the midpoints of cadences) with the fitted parameters of a geometric transit model described in Subsection \ref{ssec:fittedParameters}. The coefficients of the limb-darkening model are determined by the stellar parameters of the target star \citep{claret2011}.

The computation of the light curve is implemented in the following steps. First, an array of oversampled timestamps is constructed from the input array of timestamps. This is necessary in order to obtain an accurate flux level estimate at the temporal resolution of the data (29.4 min). For each element in the input array of timestamps, a sub-array of 11 oversampled timestamps is generated. The step size of the oversampled timestamps is 1/11 of a LC interval, or 2.67 min. The center element of the sub-array, the 6th of the 11 elements, is equal to the corresponding element in the input array of timestamps. The oversampled timestamps that fall within a given transit (including a small buffer on each side of the transit) are identified with the parameters $t_{epoch}$ and $P$. A circular Keplerian orbit, normalized by the stellar radius $R_s$, is determined from the parameters $a/R_s$ and $b$. The position vectors of the planet in the orbit are computed and the corresponding impact parameters are determined by projecting the position vectors to the plane perpendicular to the direction of the target star. The relative flux value, the ratio of the stellar flux blocked by the transiting planet to the unblocked stellar flux, is calculated for each oversampled timestamp with the impact parameter $b$, the fitted parameter $R_p/R_s$, and the limb-darkening coefficients. Finally, the relative flux at each of the input timestamps is determined as the mean of 11 relative flux values at the corresponding oversampled timestamps.

\begin{figure}
	\centering{
		\includegraphics[width=0.8\textwidth]{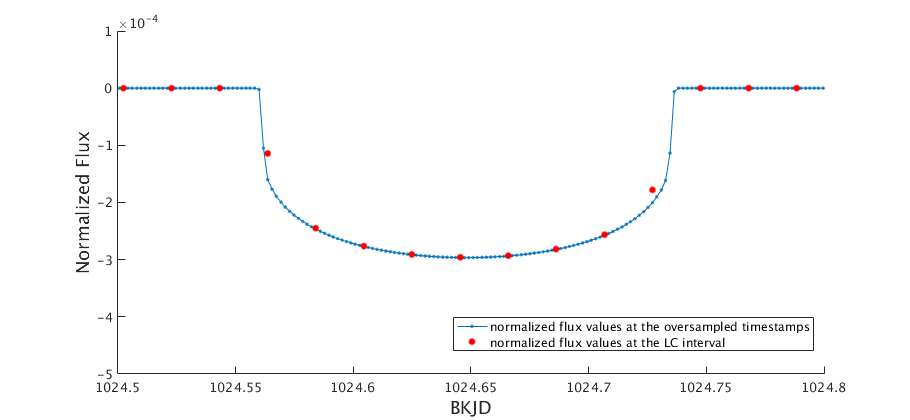}
		\caption{Normalized light curve of the 6th TCE of KIC 6541920 generated by the geometric transit signal generator. The normalized flux values at the oversampled timestamps, whose step size is 1/11 of a LC interval, or 2.67 min, are plotted in blue. Each of the normalized flux values at the LC interval, or 29.4 min, is determined as the mean of 11 corresponding values at oversampled timestamps and plotted in red.}
		\label{fig:transit-signal-generator-oversampled}}
\end{figure}

Figure \ref{fig:transit-signal-generator-oversampled} shows the normalized flux time series generated by the geometric transit signal generator with following parameters: $t_{epoch}=138.50000$ days, $P=10.30405$ days, $R_p/R_s=0.0155697$, $a/R_s=18.7471$, and $b=0.1$, which are determined by the reduced parameter fit (to be discussed in Subsection \ref{ssec:reducedParameterFits}) of the 6th TCE of the target star KIC 6541920, also known as the planet Kepler-11b. As shown in the figure, the step size of the input timestamps is the duration of a LC (29.4 min), and the normalized flux values at the input and oversampled timestamps are plotted in red and blue, respectively.

Since the surface brightness of a star appears to vary due to the limb-darkening effect, the calculation of the normalized flux value is implemented with an iterative numerical integration algorithm. At each iteration, the integration step is cut in half and an updated normalized flux time series is determined with the nonlinear limb-darkening model. The iterative process is terminated when the required precision is satisfied or when an upper limit of the execution time of the iterative algorithm is reached. If the parameter $R_p/R_s$ is less than 0.01, a small-body approximation is used to speed up the algorithm, assuming the stellar surface brightness is constant under the disc of the eclipsing object \citep{mandel2002}.

The five fitted parameters defined in Subsection \ref{ssec:fittedParameters} can be divided in two relatively independent groups: (1) the transit epoch time $t_{epoch}$ and the orbital period $P$ define the occurrence time of the transits; (2) the ratio of planet radius to star radius $R_{p}/R_{s}$, the ratio of semi-major axis to star radius $a/R_{s}$, and the impact parameter $b$, define the depth, duration, and shape of the transits. 

\begin{figure}
	\centering{
		\includegraphics[width=0.8\textwidth]{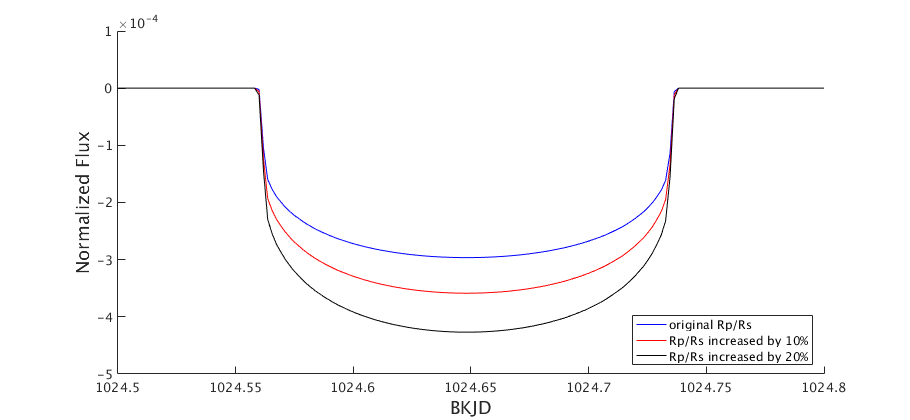}
		\includegraphics[width=0.8\textwidth]{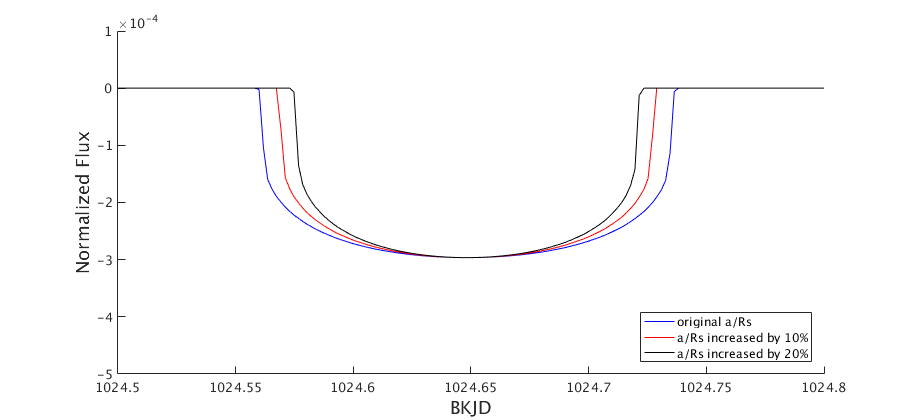}
		\includegraphics[width=0.8\textwidth]{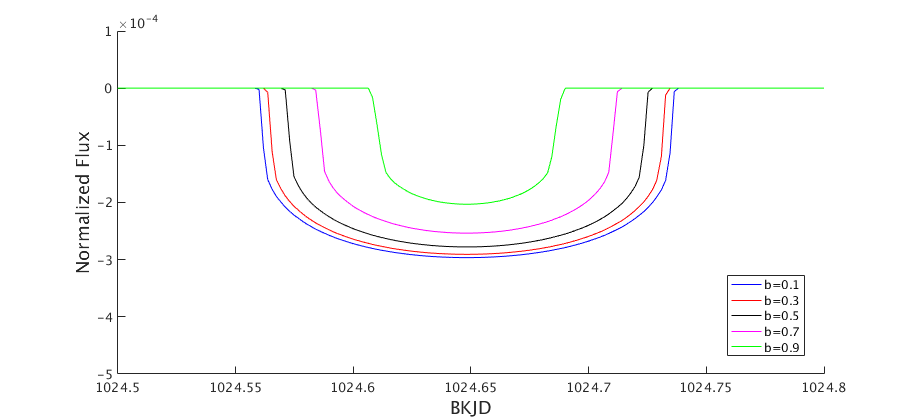}
		\caption{Light curves generated by the geometric transit signal generator with different parameters of $R_p/R_s$ (top), $a/R_s$ (middle), and $b$ (bottom). See text for an explanation.}
		\label{fig:transit-signal-generator-varying-parameters}}
\end{figure}

Figure \ref{fig:transit-signal-generator-varying-parameters} illustrates how the variations of the parameters $R_p/R_s$, $a/R_s$, and $b$ change the depth, duration, and shape of the transits in the light curves. As the reference for comparison, the light curve shown in Figure \ref{fig:transit-signal-generator-oversampled} is plotted as blue in Figure \ref{fig:transit-signal-generator-varying-parameters}. The corresponding parameter values are used as references for the parameter variations. When $R_p/R_s$ is increased by 10\% and 20\% to its reference value, the corresponding model light curves are plotted as red and black lines, respectively, in the plot on the top of Figure \ref{fig:transit-signal-generator-varying-parameters}. Since $R_p/R_s$ defines the relative size of the transiting planet to the host star, an increase of $R_p/R_s$, meaning more stellar flux is blocked by the transiting planet, leading to an increase of the transit depth. When $a/R_s$ is increased by 10\% and 20\% to its reference value, the corresponding model light curves are plotted as red and black lines, respectively, in the middle plot of Figure \ref{fig:transit-signal-generator-varying-parameters}. Since the orbital period, $P$, is fixed, an increase of $a/R_s$, indicating a decrease of the stellar radius, $R_s$, leads to a decrease of the transit duration. When $b$ is changed to 0.3, 0.5, 0.7, and 0.9, the corresponding model light curves are plotted as red, black, magenta, and green lines, respectively, in the plot on the bottom of Figure \ref{fig:transit-signal-generator-varying-parameters}. An increase of $b$ moves the transit trajectory toward the edge of the stellar disc and results in a decrease of the transit duration. Since the point at the center of the transit moves toward the edge of the stellar disc, the transit depth decreases as well due to the limb-darkening effect.   

% ***********************************************************************************************************
%
%						Geometric Model Fitting Algorithms
%
% ***********************************************************************************************************
\section{Geometric Model Fitting Algorithms}
\label{sec:GeometricModelFitting}

The inputs of the geometric model fitting algorithms include (1) the light curve after the preprocessing procedures described in Section \ref{sec:LightCurvePreprocessing}, and (2) the TCE parameters, including transit epoch time, orbital period, trial transit duration, and single and multiple event statistics, generated by the TPS component \citep{jenkins2002, jenkins2010b, jenkins2017, pt2012, pt2013, pt2014, seader2015, jdt2016}.  

Subsection \ref{ssec:iterativeWhiteningAndModelFitting} discusses an iterative whitening and model fitting process, used in the all-transit fit, odd-even transit fit, and reduced-parameter fit. The TCE parameters are used to seed the initial values of reduced parameter fits. Subsection \ref{ssec:reducedParameterFits} describes a fitting algorithm to resolve the degenerate problem of fitting the impact parameters. Subsection \ref{ssec:oddEvenTransitFit} describes the algorithms to fit odd and even transits, whose outputs are used in the diagnostic test to distinguish transiting planets from circular eclipsing binaries that have been detected at one-half of their true orbital period \citep{jdt2018}. 

The fitter outputs, which are generated when the fitting algorithm completes successfully, are described in Subsection \ref{ssec:outputs}. The alert messages, which are generated when the fitting algorithm fails, are discussed in Subsection \ref{ssec:alerts}.

\subsection{Iterative Whitening and Model Fitting}
\label{ssec:iterativeWhiteningAndModelFitting}

Compared to transit features, secular variations due to pointing drift, focus variations, and stellar variability can be quite large. Secular variations of the light curve, appearing as correlated noise, can lead to biases in the fitted parameters of the geometric transit model.  Therefore, a whitening filter is applied to the light curves before transit model fitting to account explicitly for the correlation structure of the noise. Considering that the whitening filter changes the shape of the transits, the same whitening filter is applied to the model light curve generated by the geometric transit signal generator. 

The flux time series of a target star at times $t_i$, $i=1, 2, \ldots N$, is denoted as $y(t_i)$. Let $\pmb{\theta}$ denote a 5$\times$1 vector of fitted parameters:

\begin{equation}
\pmb{\theta} = 
\left[
\begin{array}{ccccc}
t_{epoch} & P & R_p/R_s & a/R_s & b
\end{array}
\right]^T. 
\end{equation}

\noindent The predicted light curve generated from the geometric transit model with the parameter vector $\pmb{\theta}$ is denoted as $s(t_{i}, \pmb{\theta})$. When the whitening filter is applied to the time series $y(t_{i})$ and $s(t_{i}, \pmb{\theta})$, the corresponding whitened time series are denoted as $\tilde{y}(t_{i})$ and $\tilde{s}(t_{i}, \pmb{\theta})$, respectively. The geometric transit model fitting is implemented with a LM algorithm to search for the vector $\pmb{\theta}$ in the parameter space to minimize the following weighted nonlinear least-squares cost function:

\begin{equation}\label{equation:nonlinear-least-squares}
\chi^2\left(\pmb{\theta}\right) = \sum_{i=1}^{N} w_{i} \, \left[ \tilde{y}\left(t_{i}\right) - \tilde{s}\left(t_{i}, \pmb{\theta}\right) \right]^2
\end{equation}

\noindent where $w_i$, $i=1, 2, \ldots N$ are weights, ranging between 0 and 1. During the fit, these robust weights are adjusted to deemphasize points with large departures from the model values, in order to reduce the impact of outliers \citep{holland1977}.

Let $\tilde{\textbf{y}}$ and $\tilde{\textbf{s}}(\pmb{\theta})$ denote vectors of measured and predicted light curves in the whitened domain, respectively, and $\textbf{W}$ denote a diagonal matrix of the weights. Equation \ref{equation:nonlinear-least-squares} can be rewritten in the following matrix form:

\begin{equation}\label{equation:nonlinear-least-squares-matrix-form}
\chi^2(\pmb{\theta}) = [ \tilde{\textbf{y}} -  \tilde{\textbf{s}}(\pmb{\theta}) ]^T \,\textbf{W} \,[ \tilde{\textbf{y}} -  \tilde{\textbf{s}}(\pmb{\theta}) ].
\end{equation}

Since the out-of-transit light curve data just show the measurement noise around the baseline value of zero, they offer no information to characterize the transits. Therefore, the transit model fitting is restricted to the data within or close to the transits. The center times of the transits are predicted from the parameters $t_{epoch}$ and $P$, and only the light curve data whose timestamps fall in the time ranges of 5 times the transit duration, centered at the transit center, are used in the geometric transit model fit. The data selection can also be viewed as a model fit implemented with different weights to all data points;  the weight is set to 1 in Equation \ref{equation:nonlinear-least-squares} when the difference between the timestamp of the data point and the center time of the nearest transit is less than 2.5 times the transit duration; otherwise, the weight is set to 0.

\begin{figure}
	\centering{
		\includegraphics[width=\textwidth]{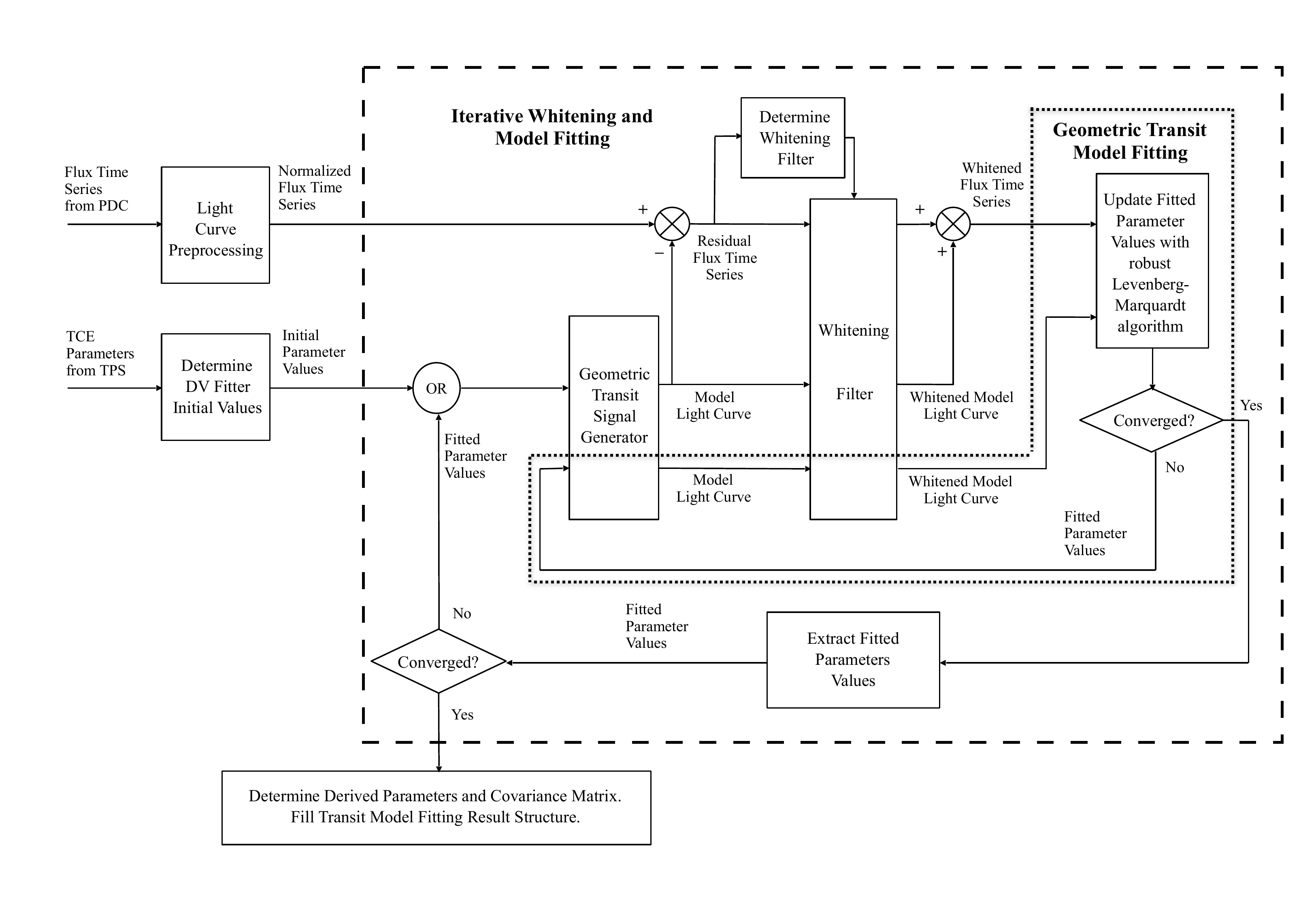}
		\caption{Block diagram of the iterative whitening and model fitting process. Two loops are shown in the figure: the outer loop, shown in the rectangle of dashed lines, includes a whitening filter and a transit model fitter, and the parameters of the whitening filter are updated on each iteration with the residuals of the transit model fitter; the inner loop, shown in the area surrounded by dotted lines, includes the LM fit and robust weight reassignment.}
		\label{fig:iterative-whitening-and-model-fitting-diagram}}
\end{figure}

For each TCE generated by TPS, the geometric transit model fitting is implemented with a loop that includes an adaptive whitening filter and a robust LM transit fitter, as shown in Figure \ref{fig:iterative-whitening-and-model-fitting-diagram}. The whitening filter transforms the correlated noise in the measured flux time series to uncorrelated, or white, noise. The predicted light curve is subjected to the same whitening filter, so the fitted parameters of the geometric transit model are determined by nonlinear least-squares fitting in the whitened domain. The fit residual is utilized to update the parameters of the whitening filter on each iteration. Robust weights are assigned to each point of the flux time series so that data with large errors are assigned small weights in the nonlinear least-squares fitting algorithm. The iterative whitening and fitting loop is terminated when both the whitening filter and the transit fitter converge or a predefined iteration limit is reached.

\subsubsection{Whitening filter}
\label{sssec:whiteningFilter}

Considering the non-stationary nature of the stellar variability, an adaptive whitening filter is generated and used to remove variations in the light curve.

The whitening filter is implemented in the following steps: (1) The flux time series and the model transit light curves are mapped into a two dimensional array of whitening coefficients, localizing the signal both in time and frequency with the Overcomplete Wavelet Transform (OWT), a modified version of the discrete wavelet transform \citep{jenkins2002, jenkins2010b, jenkins2017}; (2) The noise power in each wavelet bandpass is estimated using a moving median absolute deviation (MAD) filter; (3) The wavelet coefficients of the flux time series and the model transit light curve are normalized by the root-mean-square (rms) noise power estimates. Finally, the whitened time series is reconstructed from the updated wavelet coefficients with the inverse OWT. 

\begin{figure}
	\centering{
		\includegraphics[width=0.8\textwidth]{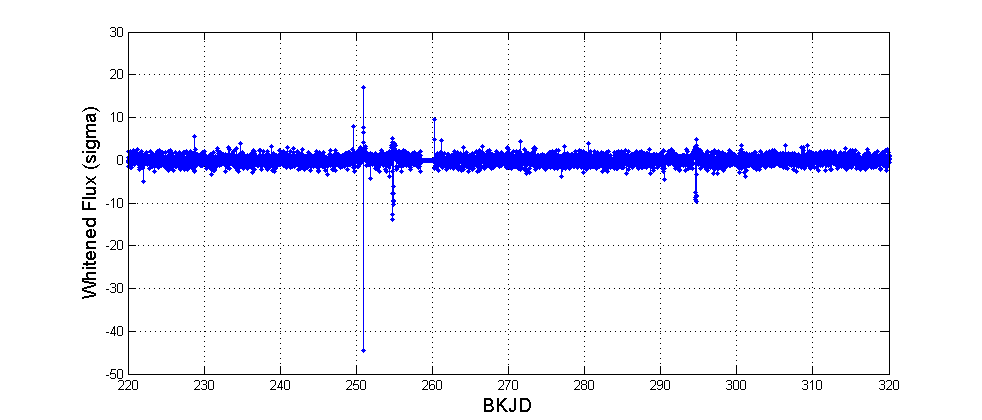}	
		\caption{Whitened flux time series of KIC 8478994.}				
		\label{fig:whitening-filter-flux-whitened}}
\end{figure}

\begin{figure}
	\centering{
		\includegraphics[width=0.8\textwidth]{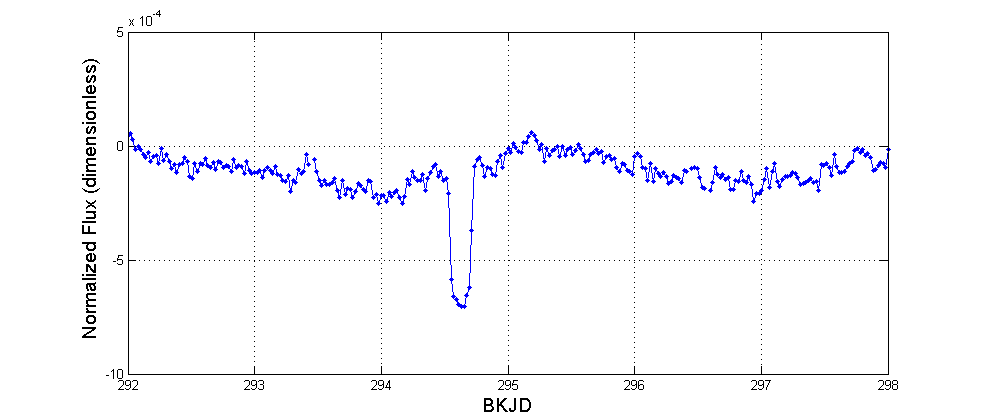}	
		\includegraphics[width=0.8\textwidth]{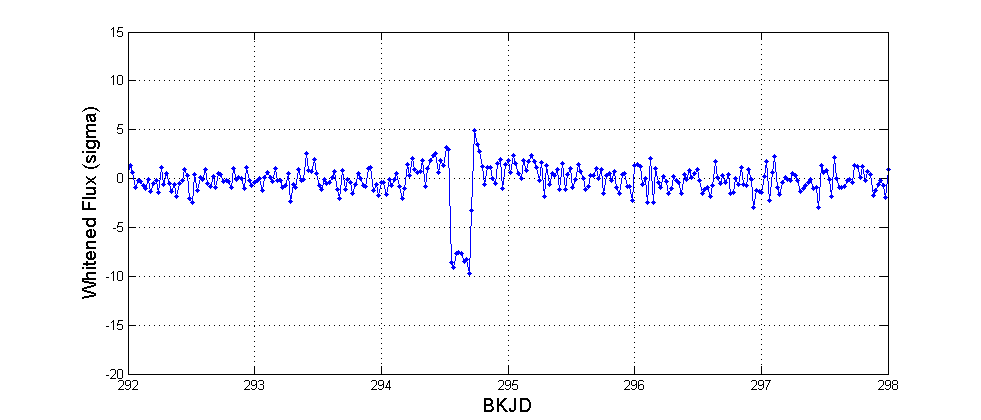}	
		\caption{Flux time series of KIC 8478994 before (top) and after (bottom) a whitening filter is applied. The length of the data segment shown in the figures is 6 days. A single transit is visible in each panel.}				
		\label{fig:whitening-filter-flux-zoomed}}
\end{figure}

Figure \ref{fig:whitening-filter-flux-whitened} shows the whitened flux time series of KIC 8478994 in an interval of 100 days, which is produced when the unwhitened normalized flux time series, shown earlier on the lower panel of Figure \ref{fig:light-curve-preprocessing-flux}, is processed with a whitening filter. Figure \ref{fig:whitening-filter-flux-zoomed} illustrates the same unwhitened and whitened flux time series in an interval of 6 days; the distortion of the whitening filter on the shape of the transit is evident. It can be seen that the depth of the transit is approximately $6\times10^{-4}$, or 600 ppm, while stellar variability produces variations of more than $3\times10^{-4}$ in the unwhitened flux time series. The whitened flux time series, whose standard deviation is equal to 1, is in units of standard deviations of the unwhitened flux.

\subsubsection{LM fit of geometric transit model parameters}
\label{sssec:LMFit}

The LM algorithm is employed to search for parameters that minimize the nonlinear least-squares cost function defined in Equation \ref{equation:nonlinear-least-squares} and \ref{equation:nonlinear-least-squares-matrix-form}.

In the general form of the LM algorithm, there is no restriction on the values of the fitted parameters. However, in the geometric transit model, the parameters $P$, $R_p/R_s$, $a/R_s$, and $b$ must be positive. Therefore, in the geometric transit model fitting algorithms, all of the fitted parameters are forced to be positive values. When an updated value of a parameter is negative in the search process, the parameter is set to the absolute value of the updated value so that all fitted parameters are positive.

An additional subtlety to the parameterization is that the impact parameter is constrained to lie in the range [0, 1] but the LM algorithm implicitly requires all fit parameters to be valid over all real values. To address this mismatch, a nonlinear transformation in the form of a $\sin$ function is performed between the ``internal'' parameter used by the LM algorithm and the ``external parameter'' used in the geometric transit model; this transformation maps the range [-$\infty$,$\infty$] used by the LM algorithm to the range [-1, 1] for the impact parameter in the geometric transit model. Negative values are also updated with the corresponding absolute values, as discussed above.

In the DR25 processing with SOC 9.3 codebase, the iterative search process for the parameter vector $\pmb{\theta}$ halted if the relative variation of the $\chi^2$ metric was less than 0.1\%, or the absolute value of the difference of the fitted parameters was less than 10\% of the corresponding uncertainties, or a preset limit of 100 iterations was reached. The threshold values are configurable DV parameters.

\subsubsection{Robust fit}
\label{sssec:robustFit}

In the weighted nonlinear least-squares fitting problem given by Equation \ref{equation:nonlinear-least-squares} or \ref{equation:nonlinear-least-squares-matrix-form}, the weight of each data point used in the fit is initialized to either 1 or 0, depending on whether the timestamp of the data point is within 2.5 times the transit duration from the center time of the nearest transit. However, when some of the selected data points are outliers, the fitting algorithm converges to a compromised solution between the valid data points and outliers, usually resulting in biases in the fitted parameters.

The robust fitting algorithm, which is optional, works by assigning a weight in the range between 0 and 1 to each data point for the fit. The outliers are assigned weights close to 0 so that the output of the robust fitting algorithm is less sensitive to the outliers in the data. The robust fitting algorithm is executed after the convergence of the non-robust LM fit. The weights are reinitialized and the LM fit is done iteratively. In each iteration, the weight of each data point is calculated from the fit residual of the previous iteration with a bisquare function, so that the data points with larger residuals are assigned smaller weights. The iterative process of weight re-assignment and LM fit continues until the fitted parameters converge within a specified tolerance.
 
\begin{figure}
	\centering{
		\includegraphics[width=0.8\textwidth]{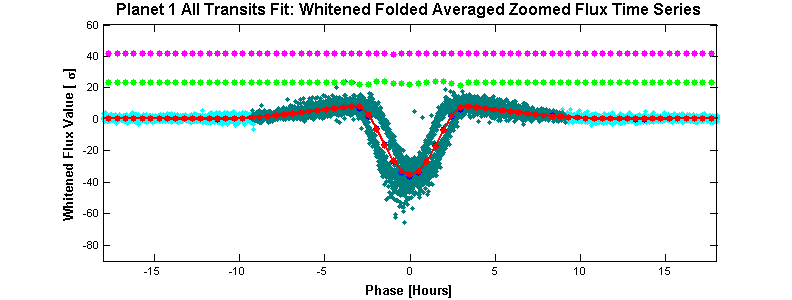}
		\includegraphics[width=0.8\textwidth]{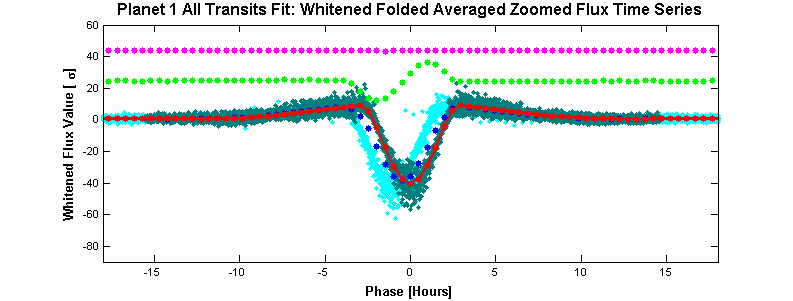}
		\caption{Folded flux time series of KIC 6960446 and model light curve generated with fitted parameters of the TCE, both in the whitened domain, when robust fit is disabled (top) and  robust fit is enabled (bottom).}\label{fig:robust-fit-flux}}	
\end{figure}

The effect of robust fit is demonstrated with the following example. The primary and secondary eclipses of an eclipsing binary target (KIC 6960446) are identified as one TCE by the TPS component. Figure \ref{fig:robust-fit-flux} shows the folded flux time series of the target and the folded model light curve generated with the fitted parameters of the TCE, both in the whitened domain, when the robust fit is off (top) and on (bottom), respectively. The secondary eclipse, which has a smaller depth, is located at phase 0 of the plot. A small phase offset is observed in the folded primary and secondary eclipses. In the plot, the flux data points are illustrated as dark green dots when the weights of the robust fit are larger than 0.1, otherwise, illustrated as light blue dots. When the weights of the fit are equally set to 1 for the data points, the fitted model light curve compromises between the primary and secondary eclipses. However, the weights of the fit are calculated iteratively in the robust fitting algorithm. As a result, most  data points of the primary eclipses are identified as outliers and assigned weights less than 0.1 at the end of the iterative process. The robust fit algorithm generates unbiased fitted parameters to characterize the secondary eclipses only.

\subsubsection{Goodness of Fit Metrics}
\label{sssec:fitGoodnessMetrics}

Two goodness of fit metrics are calculated when the transit model fitting algorithm is completed successfully. One includes the $\chi^2$ metric and the number of degrees of freedom, the other is the signal-to-noise ratio (SNR) of the fit.  

The $\chi^2$ metric is determined with Equation \ref{equation:nonlinear-least-squares}, and the number of degrees of freedom is determined as the sum of the weights minus the number of fitted parameters. It is noted the weights take values of either 0 or 1 when the robust fit is disabled, as described in Subsection \ref{sssec:robustFit}. 

The SNR of the fit is determined as:

\begin{equation}
SNR_{fit} = \sqrt{ \tilde{\textbf{s}}\left(\hat{\pmb{\theta}}\right)^T \, \textbf{W} \, \tilde{\textbf{s}}\left(\hat{\pmb{\theta}}\right) },
\end{equation}

\noindent where $\hat{\pmb{\theta}}$ is the vector of fitted parameters and $\tilde{\textbf{s}}\left(\hat{\pmb{\theta}}\right)$ is the whitened model light curve generated with $\hat{\pmb{\theta}}$. \textbf{W} is a diagonal matrix of robust weights as before.

The $\chi^2$ metric and the number of degrees of freedom measure the distance between the flux time series and the model light curve in the whitened domain. The SNR shows the strength of the TCE relative to the noise.

\subsubsection{Uncertainties of fitted and derived parameters}
\label{sssec:uncertaintiesOfParameters}

Let $\textbf{H}$ denote the Jacobian of the model light curve $\tilde{\textbf{s}}(\pmb{\theta})$ to the vector of fitted parameters $\pmb{\theta}$, such that:

\begin{equation}
\textbf{H} = \frac{\partial \, \tilde{\textbf{s}}\left(\pmb{\theta}\right)}{\partial \pmb{\theta}}.
\end{equation}

Based on the approximation to the Hessian, the covariance matrix of the fitted parameters is determined as

\begin{equation}
\mathrm{Cov}\left(\hat{\pmb{\theta}}\right) = \left(\textbf{H}^T \, \textbf{WH}\right)^{-1} \, \left(\sigma_{res}\right)^2,
\end{equation}
where $\sigma_{res}$ is the root of the mean squared (rms) value of the residuals of the fit. The elements of the Jacobian $\textbf{H}$ are determined numerically.

Let $\pmb{\alpha}$ and $\pmb{\psi}$ denote vectors of stellar parameters and derived parameters, as defined below:

\begin{equation}
\pmb{\alpha} = 
\left[
\begin{array}{ccc}
R_s & g & T_{eff}
\end{array}
\right]^T \; \mathrm{and}
\end{equation} 

\begin{equation}
\pmb{\psi} = 
\left[
\begin{array}{cccccccc}
R_p & a & i & d_{tr} & d_{in} & D & T_{eq} & \phi_{eff}
\end{array}
\right]^T ,
\end{equation}

\noindent where $g$ is the acceleration due to gravity at the surface of the star, determined from the stellar parameter $\log{g}$ as shown in Equation \ref{equation:surface-gravity}. The uncertainty of $g$ (m s$^{-2}$) can be determined from the uncertainty of $\log{g}$ ($\log_{10}$(cm s$^{-2}$)) multiplied by $g \, \ln10$.

As discussed in Subsection \ref{ssec:derivedParameters}, $\pmb{\psi}$ is a function of $\pmb{\theta}$ and $\pmb{\alpha}$. The covariance matrix of $\pmb{\psi}$, $\mathrm{Cov}\left(\pmb{\psi}\right)$, includes the components propagated both from the covariance matrix of $\pmb{\theta}$, $\mathrm{Cov}\left(\pmb{\theta}\right)$, and the uncertainties of the elements of $\pmb{\alpha}$, as shown below:

\begin{equation}
\mathrm{Cov}\left(\pmb{\psi}\right) = \left(\frac{\partial \pmb{\psi}}{\partial \pmb{\theta}}\right)^T \, \mathrm{Cov}\left(\pmb{\theta}\right) \, \left(\frac{\partial \pmb{\psi}}{\partial \pmb{\theta}}\right) + \left(\frac{\partial \pmb{\psi}}{\partial \pmb{\alpha}}\right)^T \, \mathrm{Cov}\left(\pmb{\alpha}\right) \, \left(\frac{\partial \pmb{\psi}}{\partial \pmb{\alpha}}\right)
\end{equation}

\noindent where $\mathrm{Cov}(\pmb{\alpha})$ is a diagonal matrix whose elements are squares of the uncertainties of the corresponding stellar parameters. Note that uncertainties in stellar parameters are provided by the KIC or overrides to the KIC; they are assumed to be independent. $\partial \pmb{\psi}/\partial \pmb{\theta}$ and $\partial \pmb{\psi}/\partial \pmb{\alpha}$  are Jacobians, which are described in Appendix.

The uncertainties of the fitted and derived parameters, the elements of vectors $\pmb{\theta}$ and $\pmb{\psi}$, are determined as the square roots of the diagonal elements of the matrices $\mathrm{Cov}\left(\pmb{\theta}\right)$ and $\mathrm{Cov}\left(\pmb{\psi}\right)$, respectively.

\subsection{Reduced Parameter Fits}
\label{ssec:reducedParameterFits}

Of the five fitted parameters of the geometric transit model defined in Section \ref{sec:GeometricTransitModel}, the impact parameter $b$, ranging between 0 and 1, basically describes the slope of the edges of transits. When $b$ is closer to 0, the edges are steeper. Due to the limb-darkening effect of the host star, it is difficult to determine exactly where the transit edges start and end. Therefore, in case of a low SNR for the flux time series, there is insufficient information to determine the impact parameter, which leads to large uncertainties in the fitted parameters. When DV is run with different hardware or in different computational environments, the results of the geometric transit model fit may be inconsistent, even with the same code and input data. To resolve this problem, a set of reduced parameter fits are implemented before the geometric model fitting of all transits and odd-even transits: the impact parameter $b$ is set to fixed values 0.1, 0.3, 0.5, 0.7, and 0.9, and only the parameters $t_{epoch}$, $P$, $R_p/R_s$, and $a/R_s$ are allowed to vary. After completion of the reduced-parameter fits, the all-transit fit and the odd-even transit fit follow with initial values set to the fitted parameters of the reduced-parameter fit with the minimum $\chi^2$ metric and the corresponding value of the impact parameter. 

\begin{figure}
	\centering{
		\includegraphics[width=0.8\textwidth]{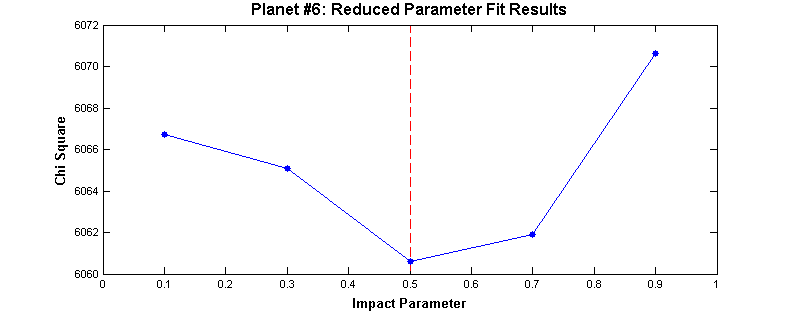}		
		\includegraphics[width=0.8\textwidth]{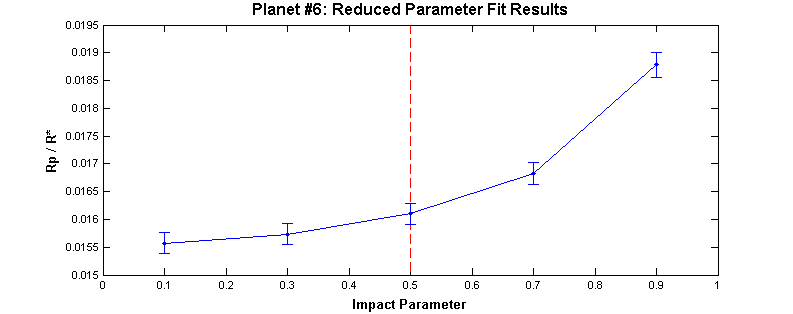}	
		\includegraphics[width=0.8\textwidth]{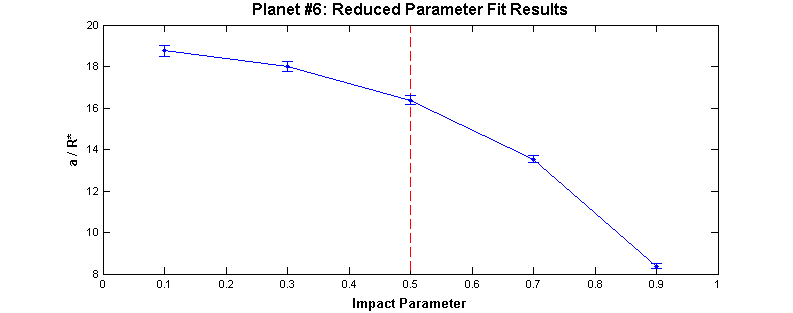}
		\caption{Reduced parameter fits of the 6th TCE of KIC 6541920: $\chi^2$ metric (top), fitted parameters $R_p/R_s$ (middle), and $a/R_s$ (bottom) vs. impact parameter, $b$.}
		\label{fig:reduced-parameter-fits-diagnostic-plots}}
\end{figure}

Figure \ref{fig:reduced-parameter-fits-diagnostic-plots} shows the diagnostic plots of the reduced parameter fits of the 6th TCE of the target star KIC 6541920. As shown in the figure, as the fixed value of $b$ increases from 0.1 to 0.9, the $\chi^2$ metric varies less than $0.2\%$ in the reduced parameter fits. However, $R_p/R_s$ increases by approximately $20\%$ and $a/R_s$ decreases by more than $50\%$. The results of the reduced parameter fit with the minimum $\chi^2$ metric are labeled with red dashed lines in the figure. As illustrated in Figure \ref{fig:transit-signal-generator-varying-parameters} of Section \ref{sec:GeometricTransitSignalGenerator}, an increase in $R_p/R_s$ leads to an increase in the transit depth, an increase in $a/R_s$ leads to a decrease in the transit duration, and an increase in $b$ results in the decrease in both the transit depth and duration. The observations of Figure \ref{fig:transit-signal-generator-varying-parameters} are consistent with the systematic variations in $R_p/R_s$ and $a/R_s$ versus $b$ in the reduced parameter fits shown in Figure \ref{fig:reduced-parameter-fits-diagnostic-plots}: when the fixed value of $b$ increases, both the transit depth and duration tend to decrease. Therefore, $R_p/R_s$ increases and $a/R_s$ decreases to compensate for the effect of the increase of $b$, so that a good fit of the model light curve to the flux time series is achieved.

\begin{figure}
	\centering{
		\includegraphics[width=0.8\textwidth]{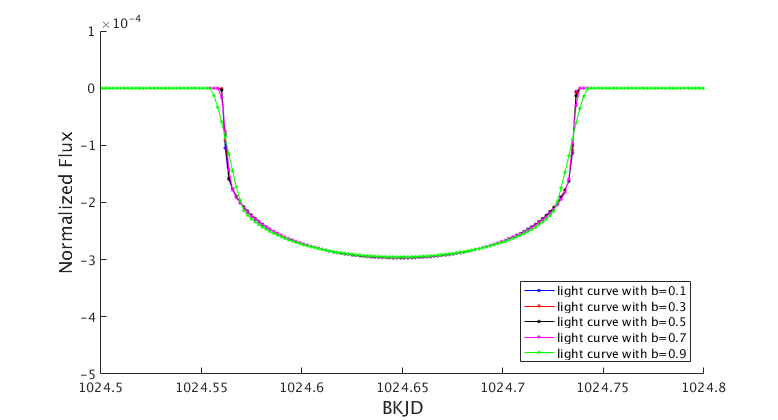}	
		\includegraphics[width=0.8\textwidth]{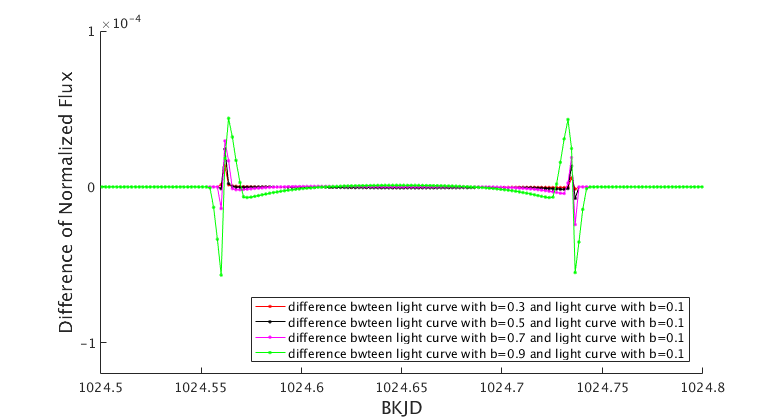}
		\caption{The plot on the top shows light curves generated with the geometric transit signal generator with the fixed $b$ values of 0.1, 0.3, 0.5, 0.7, and 0.9 and the corresponding sets of fitted parameters $t_{epoch}$, $P$, $R_p/Rs$, and $a/R_s$ of the reduced parameter fits of the 6th TCE of KIC 6541920. The plot on the bottom shows the differences between light curves with fixed $b$ values of 0.3, 0.5, 0.7, and 0.9 and one with $b=0.1$.}
		\label{fig:reduced-parameter-fits-flux}}
\end{figure}

The plot on the top of Figure \ref{fig:reduced-parameter-fits-flux} shows the light curves generated by the geometric transit signal generator with the fixed values of $b$ and the corresponding sets of fitted parameters $t_{epoch}$, $P$, $R_p/R_s$, and $a/R_s$ of the reduced parameter fits of the 6th TCE of KIC 6541920. The light curves corresponding to the fixed $b$ values of 0.1, 0.3, 0.5, 0.7, and 0.9 are plotted as blue, red, black, magenta, and green lines, respectively. The plot on the bottom of Figure \ref{fig:reduced-parameter-fits-flux} shows the differences between light curves with fixed $b$ values of 0.3, 0.5, 0.7, and 0.9 and the one with $b=0.1$.  It is observed that the difference in the light curves with different values of $b$ is small; therefore, any small variation in the input flux time series may result in a large change in the fitted parameters of $R_p/R_s$, $a/R_s$, and $b$ in the all-transit fit and odd-even transit fit.

\subsection{Odd-Even Transit Fit}
\label{ssec:oddEvenTransitFit}

When the fitting of all transits converges successfully, the same fitting algorithm is executed to fit the odd and even transits to the geometric transit model. The results of the odd-even transit fits are used in the diagnostic tests of the DV component to identify false positives generated by a circular eclipsing binary target or background eclipsing binary. 

The depths of multiple transits of a planet are ideally the same, and the transits of a planet are evenly spaced in time (in the absence of significant transit-timing variations). In contrast, the depths of primary and secondary eclipses of an eclipsing binary system are generally different due to the difference in size and brightness of the two stars. In the odd-even transit fit, two sets of parameters, one set for odd transits and the other set for even transits, are determined through an iterative whitening and model fitting process described in Subsection \ref{ssec:iterativeWhiteningAndModelFitting}, and the derived parameters are calculated for each. For each TCE, the transit depths and epochs and the corresponding uncertainties derived from the odd-even transit fit are used in the eclipsing binary discrimination tests to distinguish the flux time series of an eclipsing binary system whose primary and secondary eclipses are identified as one TCE in the TPS component. That is, the trial orbital period identified in TPS is half the true orbital period, so that the secondary eclipses are folded on top of the primary eclipses. The details of the eclipsing binary discrimination tests in the DV component are discussed in \citet{jdt2018}. 

\begin{figure}
	\centering{
		\includegraphics[width=0.8\textwidth]{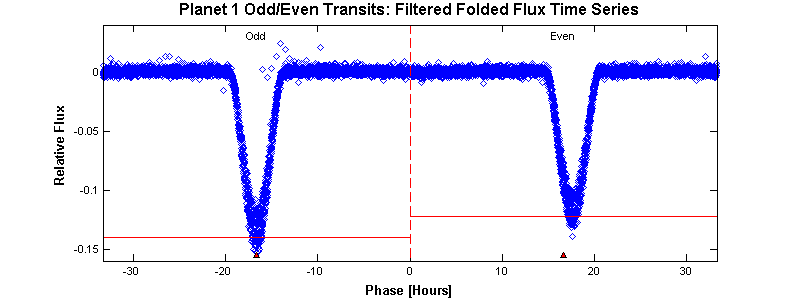}
		\caption{Folded unwhitened flux time series of the odd (left) and even (right) transits of KIC 6960446.}
		\label{fig:odd-even-transit-fit-unwhitened}}
\end{figure}

\begin{figure}
	\centering{
		\includegraphics[width=0.8\textwidth]{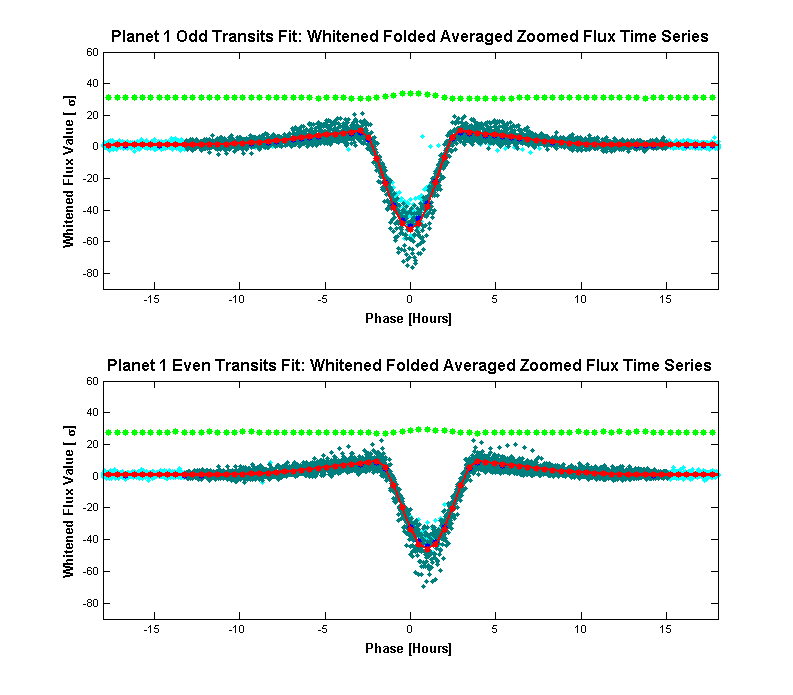}
		\caption{Folded flux time series and model light curves, both in whitened domain, of the odd (top) and even (bottom) transits of KIC 6960446.}
		\label{fig:odd-even-transit-fit-whitened}}
\end{figure}

Figure \ref{fig:odd-even-transit-fit-unwhitened} shows the folded unwhitened flux time series of the odd and even transits of the eclipsing binary target KIC 6960446. Figure \ref{fig:odd-even-transit-fit-whitened} shows the folded whitened flux time series of the odd and even transits of the same target and the folded whitened model light curves generated with fitted parameters of the odd and even transits, respectively. As shown in the figures, the primary and secondary eclipses are identified as one TCE by the TPS component, the fits of odd and even transits, which are actually primary and secondary eclipses, demonstrate that the derived transit depths of odd and even transits are different by approximately $15\%$ and that the transit epoch time of the even transits has a small offset of approximately one hour relative to that of the odd transits.

\subsection{Outputs of Geometric Transit Model Fits}
\label{ssec:outputs}

When a TCE is identified in the multiple-planet search, as described later in Section \ref{sec:MultiplePlanetSearch}, a simple check is implemented before fitting the TCE. When the eclipsing depth is more than 250,000 ppm, the TCE is labeled as a suspected eclipsing binary and geometric transit model fitting is not performed.

\begin{figure}
	\centering{
		\includegraphics[width=0.8\textwidth]{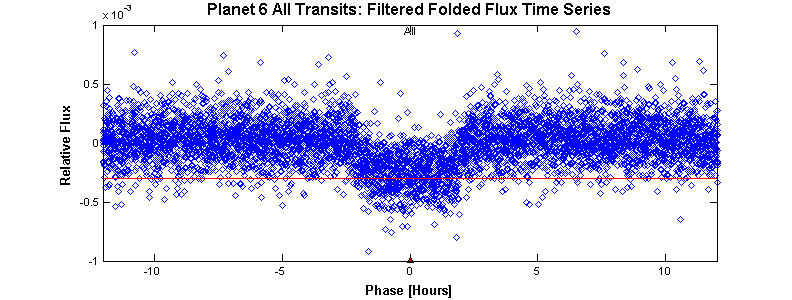}
		\includegraphics[width=0.8\textwidth]{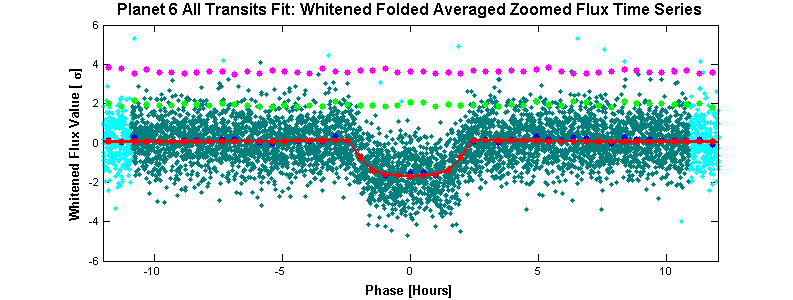}
		\caption{Folded flux time series and model light curve of the all-transit fit of the 6th TCE of KIC 6541920: unwhitened flux (top), and whitened flux and whitened model light curve (bottom).}
		\label{fig:outputs-flux}}
\end{figure}

\begin{figure}
	\centering{
		\includegraphics[width=0.8\textwidth]{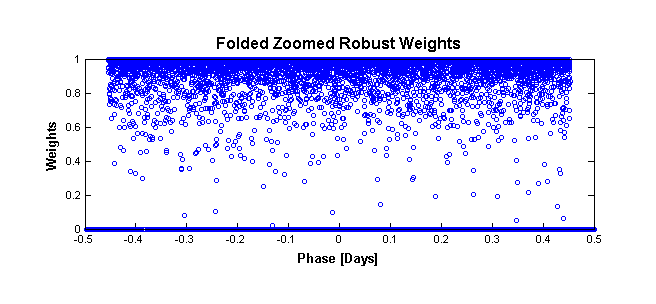}
		\caption{Folded robust weights of the all-transit fit of the 6th TCE of KIC 6541920.}
		\label{fig:outputs-robust-weights}}
\end{figure}

\begin{figure}
	\centering{
		\includegraphics[width=0.9\textwidth]{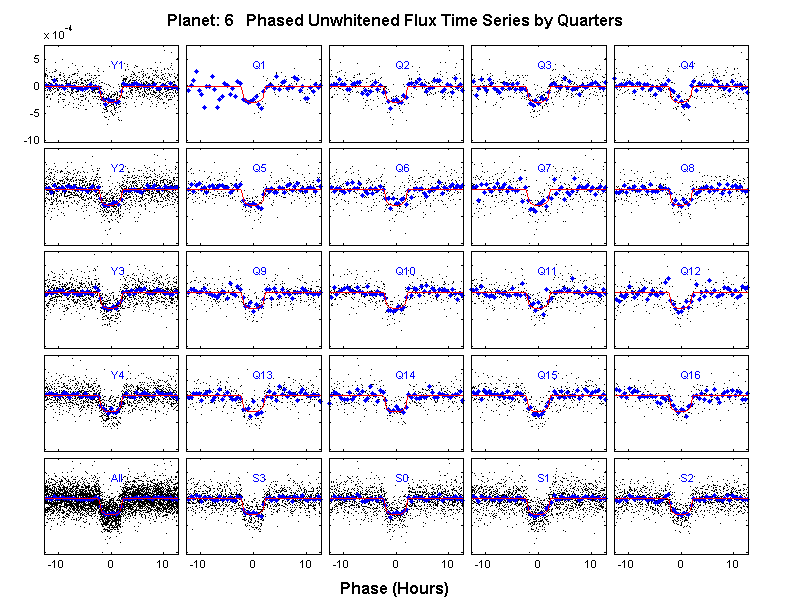}
		\caption{Folded flux time series centered on the transit events and folded model light curve of the all-transit fit, both unwhitened, of the 6th TCE of KIC 6541920 by quarter and season.}
		\label{fig:outputs-transits-by-quarter}}
\end{figure}

When the geometric transit model fitting algorithm is completed successfully, the fitted parameters and uncertainties, the derived parameters and uncertainties, and the goodness of fit metrics, etc. are saved in the DV output structure. In addition, a set of diagnostic figures are generated by the geometric transit model fitting algorithm. The diagnostic figures are included in the DV report produced for each target with at least one TCE \citep{jdt2016} and archived at the Exoplanet Archive at NExScI \citep{akeson2013}. As examples, the diagnostic plots of the all-transit fit of the 6th TCE of the target star KIC 6541920 are shown in Figure \ref{fig:outputs-flux}, Figure \ref{fig:outputs-robust-weights}, and Figure \ref{fig:outputs-transits-by-quarter}. 

The plot on the top of Figure \ref{fig:outputs-flux} shows the detrended, folded unwhitened flux time series of all transits of the TCE, and the plot on the bottom of Figure \ref{fig:outputs-flux} shows the corresponding folded whitened flux time series in the same phase range. It is noted that the vertical scales of the two plots in Figure \ref{fig:outputs-flux} are different: the unwhitened flux on the top is dimensionless while the whitened flux on the bottom is in units of the standard deviation of the unwhitened flux. The transit depth derived from the all-transit fit is illustrated with a horizontal red line in the plot on the top. In the plot on the bottom, the folded whitened light curve is illustrated in red, which is generated by the geometric transit signal generator with the fitted parameters derived from the robust fit to all transits. The flux data whose robust weights are larger than 0.1 in the all-transit fit are plotted as dark green dots, otherwise, in light blue dots. The residuals of the fit, determined as the difference of the binned average values of the whitened flux and the whitened model light curve, are plotted as green dots. The same residuals, offset by $180^{\circ}$ in phase, are plotted as magenta dots, to aid in the detection of a secondary eclipse. Figure \ref{fig:outputs-robust-weights} shows the folded weights of the robust fit of the all-transit fit of the 6th TCE of the target star KIC 6541920, in the same phase range as Figure \ref{fig:outputs-flux}.

Figure \ref{fig:outputs-transits-by-quarter} shows the detrended, folded unwhitened flux time series of the transits of the 6th TCE of the target star KIC 6541920 by quarter and season, as well as the corresponding folded unwhitened model light curves of the all-transit fit. The folded transits from the same year of the \textit{Kepler} mission are plotted in the same row, and the folded transits in the same season are plotted in the same column. For example, the folded transits in Q4 are shown in the upper right corner of the figure. The folded transits of the first year, including Q1, Q2, Q3, and Q4, are shown in the upper left corner, and the folded transits in Season 2, including Q4, Q8, Q12, and Q16, are shown in the lower right corner. At the lower left corner, the folded transits in all 17 quarters of the \textit{Kepler} science data are illustrated. 

For the odd-even transit fit, as illustrated in Figure \ref{fig:odd-even-transit-fit-unwhitened} and Figure \ref{fig:odd-even-transit-fit-whitened}, the plots of folded unwhitened flux time series of odd and even transits are placed horizontally, and the plots of folded flux time series and folded model light curves, both whitened, are placed vertically, so that the difference in the derived depths and the offset in the fitted transit epoch times of the odd and even transits can be easily observed.

For the reduced parameter fits, a set of diagnostic plots, the same as those of the all-transit fit, are generated for each fit. In addition, as illustrated in Figure \ref{fig:reduced-parameter-fits-diagnostic-plots}, several diagnostic plots are generated to illustrate the variations of the $\chi^2$ metric and fitted parameters versus the fixed value of the impact parameter. The fit with the minimum $\chi^2$ metric is labeled with red dash lines on these figures.

\subsection{Alerts of Failed Fits}
\label{ssec:alerts}

When the geometric transit model fit fails, an alert is generated indicating the nature of the failure and where it occurs. These alerts are included in the Appendix of the DV report. Table \ref{table:alerts} lists the top five alerts of the failed all-transit fits in the DR25 DV processing.

As shown the the table, the most common failure of the all-transit fits is that the time used by the fitting algorithm goes beyond the preset limit and the fit is stopped during the call of the function ``model\_function.'' This usually happens when an anomalously noisy flux time series is fitted with a transit model; the criterion of convergence can never be met and the algorithm goes into an infinite loop.

In the fitting algorithm, several check points are set to verify the validity of the fit results, or else the fit results are labeled as invalid. For example, the fitted parameter of the transit epoch time $t_{epoch}$ should fall in a range centered on the corresponding TCE value given by the TPS component. Furthermore,  the derived transit duration cannot be smaller than the duration of a LC interval (29.4 min). As shown in items 2 and 3 of Table \ref{table:alerts}, the alerts of invalid fit results are generated during the call of the functions ``fit\_transit'' and ``fill\_planet\_results.''

The fourth alert of Table \ref{table:alerts} is generated when the time used in the iterative numerical integration algorithm, as described in Section \ref{sec:GeometricTransitSignalGenerator}, exceeds the preset limit in the call of the function to compute the transit light curve when the small-body approximation is not applicable. The fifth alert occurs during the call of the function ``transitFitClass'' when too many flux data points are gapped and the number of remaining transits is less than 2 in the all-transit fit; as a result, there is insufficient information to determine reliable parameters of the transit model.

\begin{table}
	\centering{
		\caption{Top five alerts of failed all-transit fits in DR25 DV run.}
		\begin{tabular}[width=.9\textwidth]{ccc}
			\textbf{Index} & \textbf{Alert Type}                                       & \textbf{Number} \\
			\hline 
			1              & dv:modelFunction:fitTimeLimitExceeded                     & 1,012           \\
			2              & dv:fitTransit:transitEpochBkjdBigDifferenceFromTceValue   & 592             \\
			3              & dv:fillPlanetResults:transitDurationSmallerThanLowerBound & 262             \\
			4              & dv:computeLargeBodyTransitLightCurve:takingTooLong        & 45              \\     
			5              & dv:transitFitClass:insufficientTransitsToFit              & 41              \\
		\end{tabular}
		\label{table:alerts}}
\end{table}

% ***********************************************************************************************************
%
%						Trapezoidal Model Fitting Algorithm
%
% ***********************************************************************************************************
\section{Trapezoidal Model Fitting Algorithm}
\label{sec:TrapezoidalModelFitting}

As an optional configuration of the transit model fitting in the DV component, the light curve of the target for which a TCE is generated can also be fitted by a trapezoidal model. The trapezoidal model is a simple description of the basic characteristics of the transits, and may converge to a successful fit when the limb-darkened transit model fit fails. In these cases, the trapezoidal model fit parameters can be used to support subsequent DV diagnostic tests, which otherwise could not be performed \citep{jdt2018}.

The trapezoidal model includes the following four fitted parameters: 

\begin{itemize}
	\item Transit epoch time $t_{epoch}$ (BKJD): same as the fitted parameter of the geometric model defined in Subsection \ref{ssec:fittedParameters};
	\item Transit depth $D$ (ppm): same as the derived parameter of the geometric model defined in Subsection \ref{ssec:derivedParameters};	
	\item Transit mid-duration $d_{mid}$ (hours): the duration of transit at half of the transit depth, as illustrated in Figure \ref{fig:transit-model-diagram};
	\item  Ratio of ingress time to mid-duration $d_{in}/d_{mid}$ (dimensionless): the transit ingress time $d_{in}$ is same as the derived parameter defined in Subsection \ref{ssec:derivedParameters},  but this is the ratio of the ingress time to mid-duration.
\end{itemize}

The orbital period $P$ (days) is set to the corresponding TCE parameter value provided by the TPS component in the transit signal generator with the trapezoidal model.

An alternative detrending algorithm based on the nonparametric penalized least squares method from \citet{garcia2010} is applied to the PDC light curve prior to the trapezoidal model fit. The algorithm allows for missing data via weight assignment and solves for the free parameter controlling the amount of smoothing using a generalized cross validation method. To prevent suppression of the transit signal we treat data in transit according to the TCE ephemeris and transit duration as missing with a weight of zero. Each quarterly PDC light curve is detrended independently.  When a high frequency (similar or shorter time scale than the transit signal) astrophysical signal is present in a light curve, the automated method for determining the smoothing parameter results in unwanted suppression of the transit signal.  To prevent over-smoothing, the smoothing parameter is determined on a light curve with a low-pass filter applied.  The low-pass filtered light curve is generated by subtraction of a high-pass (simple two-point difference) filtered version of the light curve.  The adopted detrending model, which results from using the smoothing parameter estimated from the low-pass filtered version of the light curve, is used in normalization of the PDC light curve.

The trapezoidal model fitting algorithm is implemented with 10 repeated LM fits. For each fit, the initial value of the fitted parameter is set randomly with a uniform distribution in a pre-determined range. The outputs of the trapezoidal model fitting algorithm are determined as those of the LM fit with the minimum $\chi^2$ metric. 

\begin{figure}
	\centering{
		\includegraphics[width=0.8\textwidth]{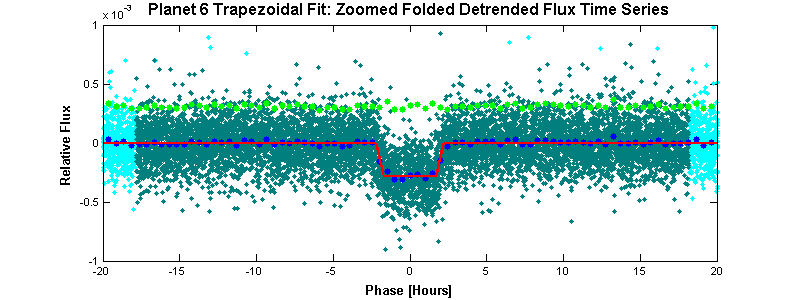}
		\caption{Folded flux time series and folded model light curve of the trapezoidal model fit, both unwhitened, of the 6th TCE of KIC 6541920.}
		\label{fig:trapezoidal-model-fitting-unwhitened}}
\end{figure}

Figure \ref{fig:trapezoidal-model-fitting-unwhitened} shows a diagnostic plot generated in the trapezoidal model fit of the 6th TCE of KIC 6541920. Only the flux data whose timestamps fall in the time ranges of 8 times the transit duration (one of the TCE parameters generated by the TPS component) and centered at the transit center time are employed in the trapezoidal model fit. The flux data points within this range used in the fit are plotted as dark green dots in the figure, otherwise, in light blue dots. The folded light curve generated by the trapezoidal model with the fitted parameters is plotted as red lines and the residual of the fit is offset vertically for clarity and plotted as green dots. Since the whitening filter, described in Subsection \ref{sssec:whiteningFilter}, is not used in the trapezoidal model fitting algorithm, all the data shown in Figure \ref{fig:trapezoidal-model-fitting-unwhitened} are in the unwhitened domain.

Compared to the plot on the bottom of Figure \ref{fig:outputs-flux} of Subsection \ref{ssec:outputs}, the bottom of the transit is flat in the model light curve shown in
Figure \ref{fig:trapezoidal-model-fitting-unwhitened} since the limb-darkening effect is not included in the trapezoidal transit model.

The trapezoidal model fit provides a quick assessment of the transit signal. The fitted trapezoidal transit model is used in the diagnostic tests of the DV component when the fit with the geometric transit model fails or when the fit is not performed, such as for suspected eclipsing binaries. 

% ***********************************************************************************************************
%
%						Multiple Planet Search
%
% ***********************************************************************************************************
\section{Multiple-Planet Search}
\label{sec:MultiplePlanetSearch}

After the fitting process has completed, the data points within 1.5 times the transit duration from the central time of the nearest transit are removed, where the transit duration and the central time of transits are determined from the fitted parameters of the all-transit fit. So the signature of the known TCE is removed, and the residual flux is subjected to a search for additional planets by calling TPS in the DV component. The transit model fitting algorithms, including the reduced parameter fits, all-transit fit, odd-even transit fit, and the trapezoidal model fit, are applied again if an additional TCE is generated. The search for additional planets concludes when no additional TCEs are produced or an iteration limit is reached, as shown in the flowchart of Figure \ref{fig:flowchart-transit-model-fitting-and-multiple-planet-search}.  

\begin{figure}
	\centering{
		\includegraphics[width=0.9\textwidth]{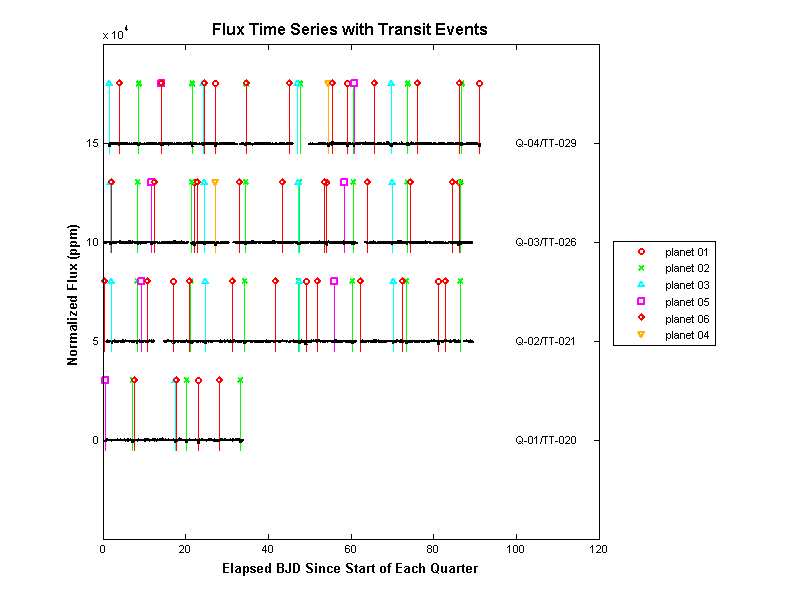}
		\caption{Light curve of KIC 6541920 from Q1 to Q4 and transits of six TCEs.}
		\label{fig:multiple-planet-search-tce-transits}}
\end{figure}

Figure \ref{fig:multiple-planet-search-tce-transits} shows the light curve of KIC 6541920 (Kepler-11) from Q1 to Q4. The quarterly segments are offset vertically for clarity. The transits of six TCEs are labeled with different colors and symbols in the figure. The first TCE, labeled with red circles, is identified by the TPS component and the corresponding parameters to characterize the TCE are provided to DV. The remaining five TCEs are identified in the multiple-planet search by calling TPS directly in the DV component.

\begin{figure}
	\centering{
		\includegraphics[width=0.8\textwidth]{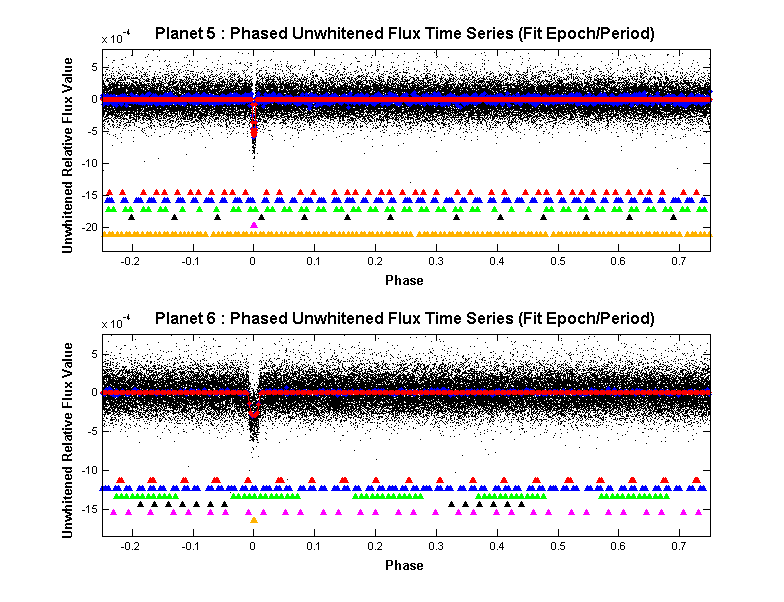}
		\caption{Phased flux time series of KIC 6541920 with the fitted parameters $t_{epoch}$ and $P$ of the 5th (top) and 6th (bottom) TCEs, respectively.}
		\label{fig:multiple-planet-search-phased-flux}}
\end{figure}

Figure \ref{fig:multiple-planet-search-phased-flux} shows the folded flux time series of KIC 6541920 in the unwhitened domain, phased with the fitted parameters $t_{epoch}$ and $P$ of the 5th and 6th TCEs, respectively. The binned average values of the folded flux and the folded model light curve are plotted as blue and red dots, respectively. The triangles in different colors show the location of the transits of all six TCEs in the phased flux time series.

% ***********************************************************************************************************
%
%						Performance of Transit Model Fitting and Multiple-Planet Search
%
% ***********************************************************************************************************
\section{Performance of Transit Model Fitting and Multiple-Planet Search}
\label{sec:PerformanceOfTransitModelFittingAndMultiplePlanetSearch}

The  17 quarters of primary mission science data, collected by the \textit{Kepler} spacecraft from May 13, 2009 to April 8, 2013, were processed by the SOC 9.3 codebase of the \textit{Kepler} Data Processing Pipeline in January 2016. 17,230 target stars, which generated TCEs in the TPS component, were processed successfully by the DV component. This pipeline run is referred to as DR25, and the TCE population was described in \citet{jdt2016}. 

Among a total of 34,032 TCEs generated in the TPS component and in the multiple-planet search of the DV component, 239 ($0.7\%$) TCEs were labeled as suspected eclipsing binaries, 2,062 ($6.1\%$) TCEs failed in the all-transit fit, and 31,731 ($93.2\%$) TCEs completed the all-transit fit successfully. Out of 31,731 TCEs with successful all-transit fits, 2,620 ($8.3\%$) TCEs failed in the odd-even transit fit, and 29,111 ($91.7\%$) TCEs completed the odd-even transit fit successfully. 33,125 ($97.3\%$) out of 34,032 TCEs completed the trapezoidal model fit successfully.

\begin{figure}
	\centering{
		\includegraphics[width=0.45\textwidth]{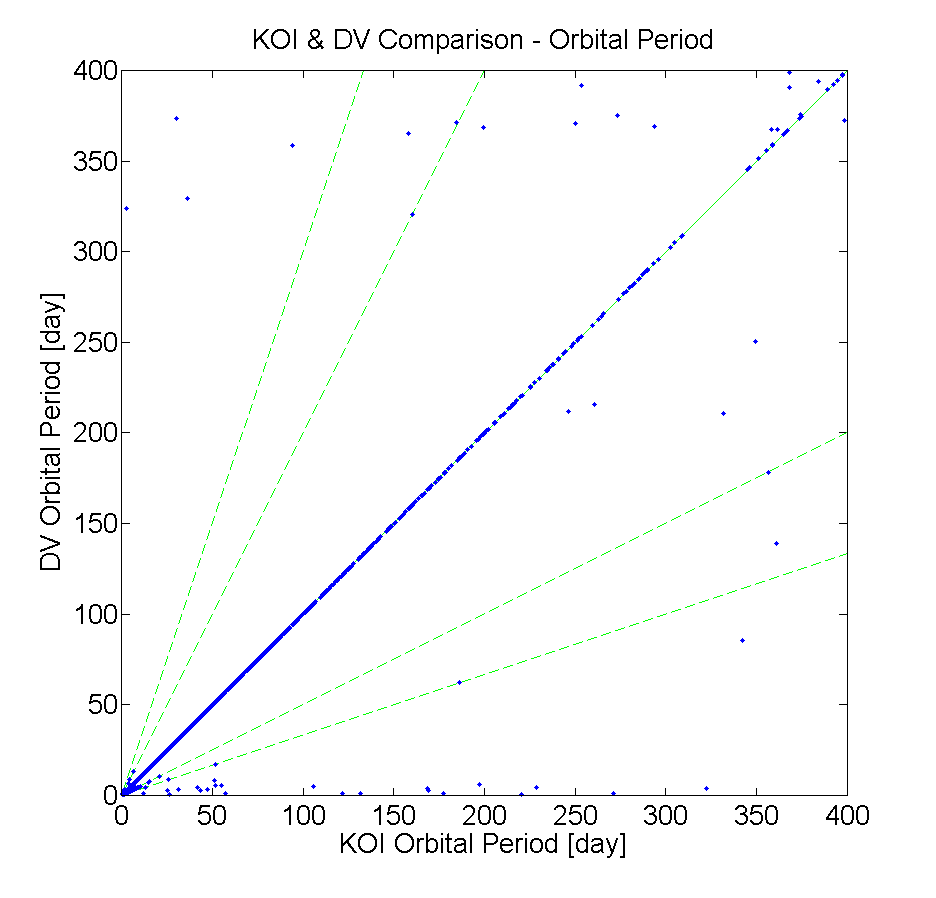}
		\includegraphics[width=0.45\textwidth]{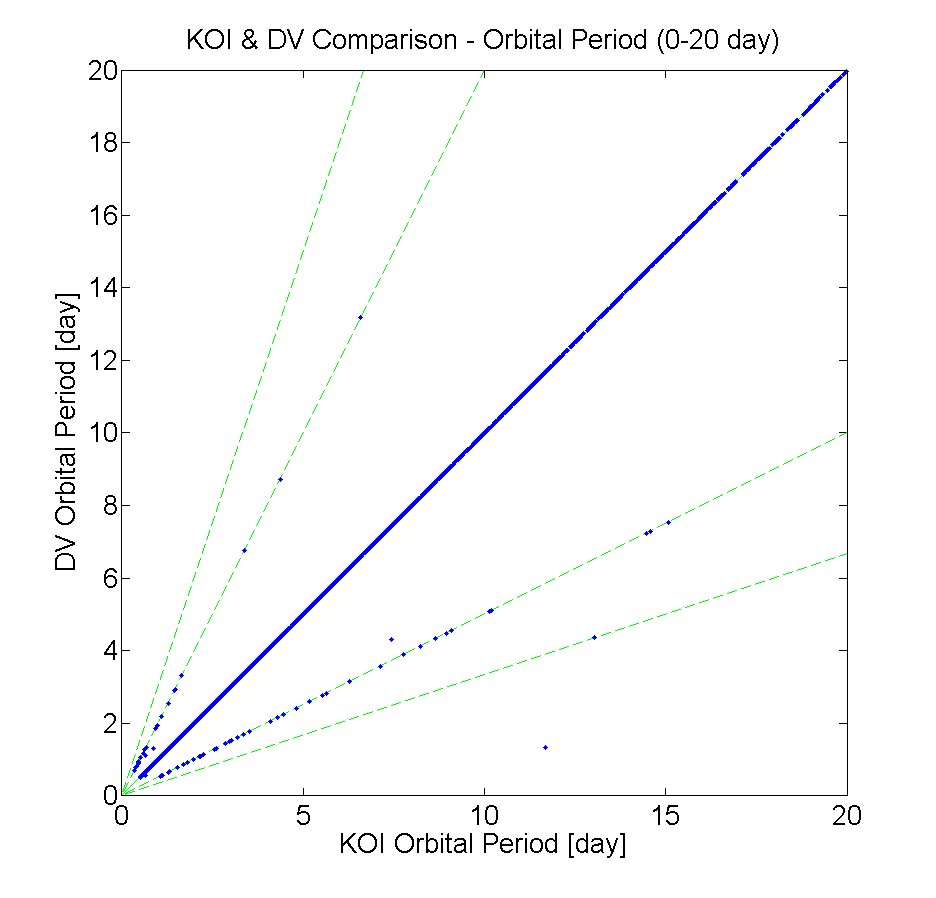}
		\caption{Comparison of DV Fitted parameters and KOI parameters: all orbital periods (left) and orbital periods ranging from 0 to 20 days (right).}
		\label{fig:performance-compare-period}}
\end{figure}

Figure \ref{fig:performance-compare-period} compares the orbital period of the DV all-transit fit and the corresponding KOI parameter produced independently \citep{rowe2014}. The plot on the left shows all orbital periods in the comparison and the plot on the right shows the orbital periods  ranging from 0 to 20 days only. The diagonal green line shows where the DV fitted orbital period value is equal to the KOI parameter value; the other four green lines indicate that the two period values differ by a factor of $1/3$, $1/2$, 2, and 3, respectively. It is observed in Figure \ref{fig:performance-compare-period} that the orbital periods of some TCEs identified in TPS and DV are double or half of the corresponding KOI values.

\begin{figure}
	\centering{
		\includegraphics[width=0.45\textwidth]{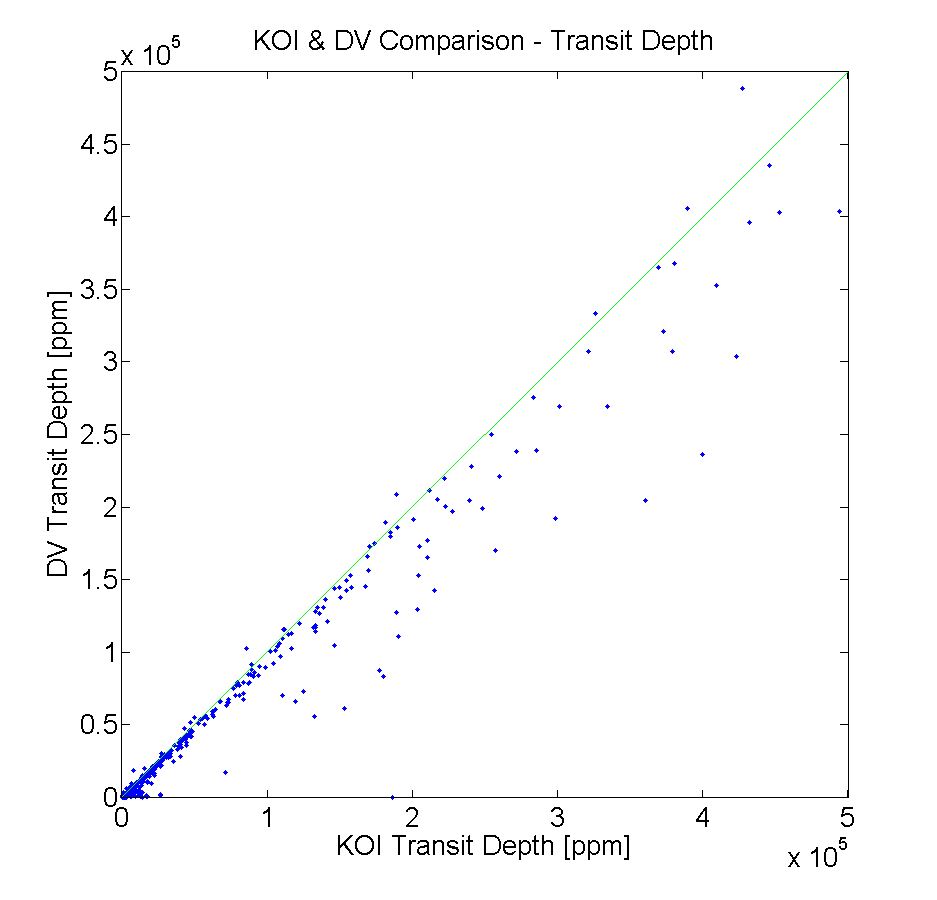}
		\includegraphics[width=0.45\textwidth]{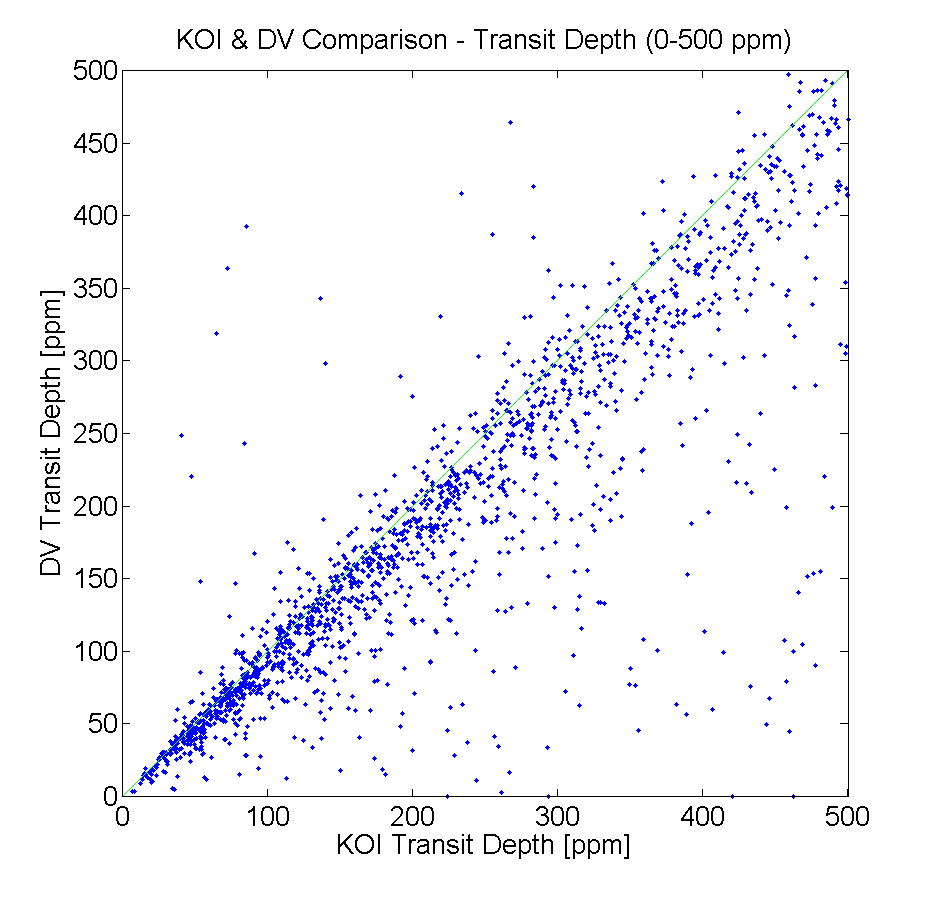}
		\caption{Comparison of DV Fitted parameters and KOI parameters: all transit depths (left) and transit depths ranging from 0 to 500 ppm (right).}
		\label{fig:performance-compare-depth}}
\end{figure}

Figure \ref{fig:performance-compare-depth} compares the transit depth derived from the DV all-transit fit and the corresponding KOI parameter. Similar to Figure \ref{fig:performance-compare-period}, the plot on the left shows all-transit depths in the comparison and the plot on the right shows the transit depths ranging from 0 to 500 ppm only. The diagonal green line shows where the DV fitted transit depth value is equal to the KOI parameter value. It is observed that the KOI values of the transit depth are larger than the corresponding DV fitted values for many TCEs. Investigations show some short-period transit signals are degraded in the light curve preprocessing procedure of harmonic removal when the orbital period is small \citep{christiansen2013, christiansen2015}.

\begin{figure}
	\centering{
		\includegraphics[width=0.7\textwidth]{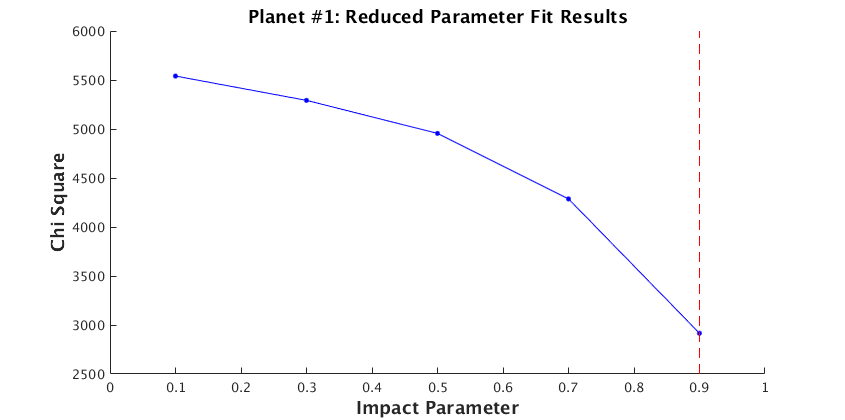}		
		\includegraphics[width=0.7\textwidth]{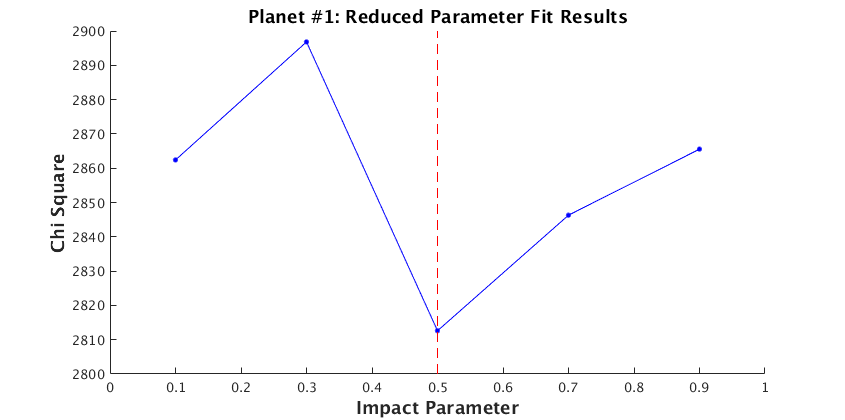}			
		\caption{The diagnostic plots of  $\chi^2$  versus $b$ of the reduced parameter fits of the 1st TCE of KIC 6541920, generated by the SOC 9.3 codebase in January 2016 (top) and the modified SOC 9.3 codebase in August 2016 (bottom), respectively. As shown in the plot on the top, due to a software defect introduced into the 9.3 codebase, the $\chi^2$ metric of the reduced parameter fit systematically decreases as the fixed value of the impact parameter $b$ increases. In the plot on the bottom, there is no systematic decrease of the $\chi^2$ metric as $b$ increases.}
		\label{fig:performance-reduced-parameter-fits-chi-square}}
\end{figure}

A software defect introduced into the SOC 9.3 code for the reduced parameter fits came to light after the DR25 run. As discussed in Subsection \ref{ssec:iterativeWhiteningAndModelFitting}, only the data points within the range of the transit and a buffer on each side of the transit are employed in the weighted nonlinear least-squares fitting. The weights are assigned 1 and 0, respectively, depending on whether the data points are used in the fitting or not. As shown in Equation \ref{equation:nonlinear-least-squares}, the $\chi^2$ metric is related to how many data points are used in the fit: the more data points used in the fit, the larger the $\chi^2$ metric. In the SOC 9.3 codebase, the data points employed in the reduced parameter fits are related to the fixed value of the impact parameter $b$. As a result, the calculated $\chi^2$ metric is improperly related to the value of $b$: the closer $b$ is to 1, the smaller the $\chi^2$ metric. The software defect was corrected in a modified SOC 9.3 codebase, which was used in a supplemental DV run in August 2016. Figure \ref{fig:performance-reduced-parameter-fits-chi-square} shows the diagnostic plots of the $\chi^2$ metric versus $b$ of the reduced parameter fits of the 1st TCE of KIC 6541920 (the planet Kepler-11e), which were generated by the SOC 9.3 codebase in January 2016 and the modified SOC 9.3 codebase in August 2016, respectively. As shown in the plot on the top of Figure \ref{fig:performance-reduced-parameter-fits-chi-square}, due to the software defect, the $\chi^2$ metric systematically decreases as $b$ increases so the result of the reduced parameter fit with the fixed value of $b=0.9$ is always selected to seed the all-transit fit. The same was true for all TCEs in the DR25 DV run. In the plot on the bottom of Figure \ref{fig:performance-reduced-parameter-fits-chi-square}, there is no systematic decrease of the $\chi^2$ metric as $b$ increases, and $b=0.5$ is selected to seed the all-transit fit.

\begin{figure}
	\centering{
		\includegraphics[width=0.7\textwidth]{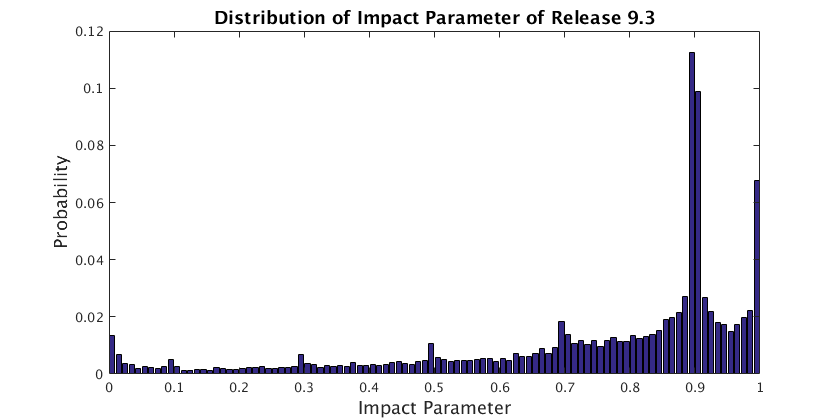}		
		\includegraphics[width=0.7\textwidth]{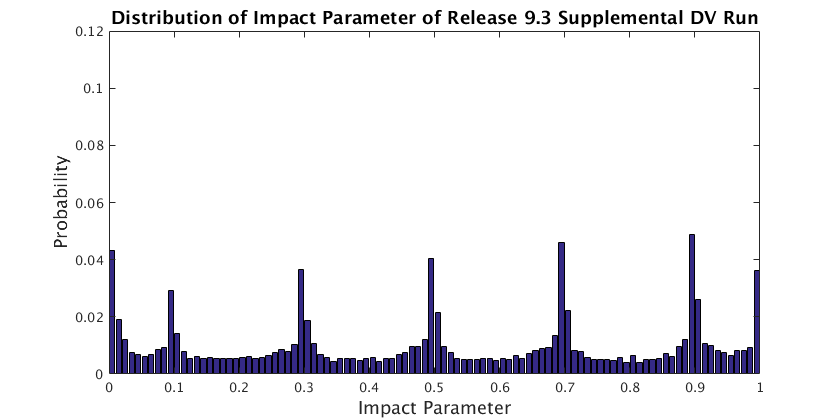}	
		\caption{Distribution of the fitted parameter $b$ of the all-transit fits of a set of 16,514 TCEs, generated by the SOC 9.3 codebase in January 2016 (top) and the modified SOC 9.3 codebase in August 2016 (bottom), respectively. As shown in the plot on the top, the distribution of the fitted parameter $b$ is biased toward $b=0.9$ in the outputs of the all-transit fit of the SOC 9.3 codebase. In the plot on the bottom, there is no bias toward $b=0.9$ in the distribution of the fitted parameter $b$ in the outputs of the all-transit fit of the modified SOC 9.3 codebase.}
		\label{fig:performance-impact-parameter-distribution-all}}
\end{figure}

\begin{figure}
	\centering{
		\includegraphics[width=0.7\textwidth]{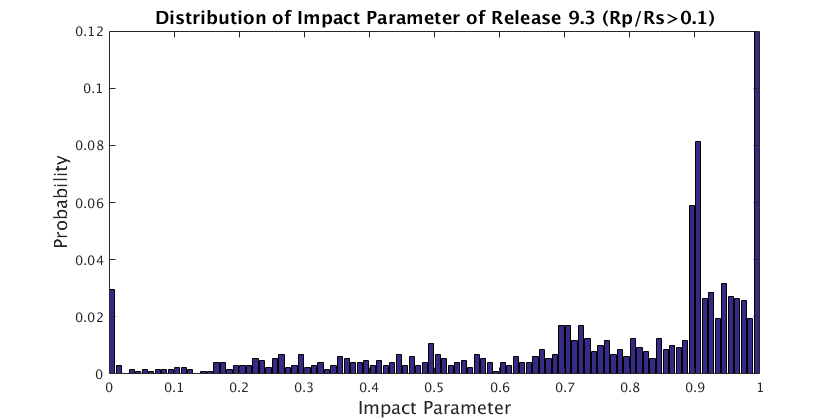}		
		\includegraphics[width=0.7\textwidth]{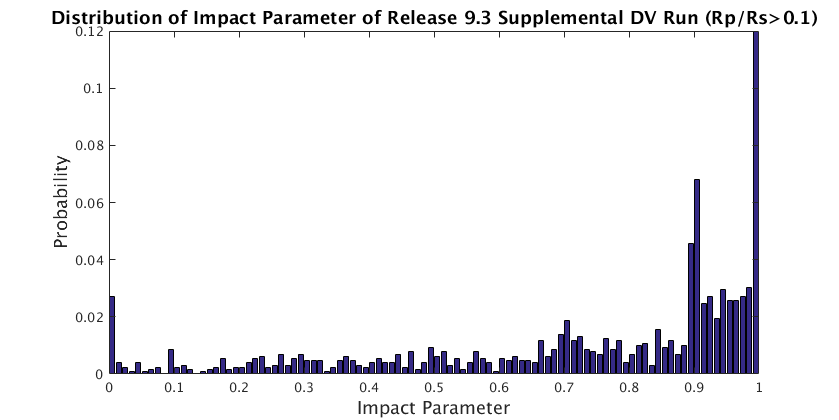}	
		\caption{Distribution of the fitted parameter $b$ of the all-transit fits of a set of 1,292 TCEs, generated by the SOC 9.3 codebase in January 2016 (top) and the modified SOC 9.3 codebase in August 2016 (bottom), respectively. The set of 1,292 TCEs, a subset of the 16,514 TCEs, was selected as the fitted parameter $R_p/R_s$ was larger than 0.1 in the supplemetal DV run in August 2016. It is observed that the convergence of the all-transit fit is essentially independent of the initial seed value of the impact parameter $b$ for large planets.}
		\label{fig:performance-impact-parameter-distribution-large}}
\end{figure}

As shown in Subsection \ref{ssec:reducedParameterFits}, flux time series with low SNR including those with transiting planet signatures of small planets (relative to the size of their host stars) may be well fitted over a wide range of impact parameter values. Figure \ref{fig:performance-impact-parameter-distribution-all} shows the distributions of the fitted parameter $b$ in the all-transit fits of a set of 16,514 TCEs, generated in the DR25 run with the SOC 9.3 codebases in January 2016 and in the supplemental DV run with the modified SOC 9.3 codebase in August 2016, respectively. The 16,514 TCEs were selected from the 1st TCEs of the targets, which completed the all-transit fits successfully in both runs. The distribution of the fitted parameter $b$ is biased toward the initial seed value of $b=0.9$ in the outputs of the all-transit fits with the SOC 9.3 codebase, as shown in the plot on the top of Figure \ref{fig:performance-impact-parameter-distribution-all}. In the plot on the bottom of Figure \ref{fig:performance-impact-parameter-distribution-all}, there  is no bias toward $b=0.9$ in the distribution of the fitted parameter $b$ in the all-transit fits with the modified SOC 9.3 codebase. Figure \ref{fig:performance-impact-parameter-distribution-large} shows the distributions of the fitted parameter $b$ in the all-transit fits of a set of 1,292 TCEs in both DV runs. The set of 1,292 TCEs, a subset of the 16,514 TCEs, was selected as the fitted parameter $R_p/R_s$ was larger than 0.1 in the supplemetal DV run in August 2016. It is observed that the convergence of the all-transit fit is essentially independent of the initial seed value of the impact parameter $b$ for large planets. 

As discussed by \citet{jdt2016}, transiting planets with a high impact parameter must be larger than those with a lower impact parameter for given transit depths on the same host stars because of the limb-darkening effect. It is noted that all planetary candidates in the DR25 \textit{Kepler Mission} catalog by \citet{thompson2018} were modeled independently by the TCE Review Team (TCERT), so the bias discussed here relates only to TCE products of the SOC 9.3 DR25 of the \textit{Kepler} Science Data Processing Pipeline at the NASA Exoplanet Archive.

% ***********************************************************************************************************
%
%                   	Conclusions
%
% ***********************************************************************************************************
\section{Conclusions}
\label{sec:Conclusions}

We have presented the transit model fitting and multiple-planet search algorithm of the Data Validation component of the \textit{Kepler} Science Data Processing Pipeline. The performance of the algorithm is demonstrated by the results of processing 17 quarters of \textit{Kepler} science data using SOC 9.3 codebase of the \textit{Kepler} Science Data Processing Pipeline in January 2016 (DR25). The results of the transit model fitting of the TCEs identified by the pipeline are accessible by the science community at the NASA Exoplanet Archive. The \textit{Kepler} SOC 9.3 codebase is also available to the general public through GitHub. A software defect that biased the seeding of the limb-darkened model fits and ultimately the model fit results for small planets was corrected in a modified SOC 9.3 codebase, which was implemented in a supplemental DV run after DR25.

% ***********************************************************************************************************

\acknowledgements{Funding for the \textit{Kepler Mission} was provided by the NASA Science Mission Directorate. The data validation products were generated by the \textit{Kepler} Science Data Processing Pipeline through the efforts of the \textit{Kepler} Science Operations Center and Science Office at NASA Ames Research Center.}

{\it Facilities:} \facility{Kepler}

\appendix
\section{Jacobians in Subsection \ref{sssec:uncertaintiesOfParameters}}

The Jacobians $\partial \pmb{\psi}/\partial \pmb{\theta}$ and $\partial \pmb{\psi}/\partial \pmb{\alpha}$  in Subsection \ref{sssec:uncertaintiesOfParameters} have the following forms:

\begin{equation}
\frac{\partial \pmb{\psi}}{\partial \pmb{\theta}} = 
\left[
\begin{array}{ccccc}
0 & 0                                      & \frac{\partial    R_p}{\partial \left(R_p/R_s\right)} & 0                                                   & 0                                  \\
0 & \frac{\partial          a}{\partial P} & 0                                                     & 0                                                   & 0                                  \\
0 & 0                                      & 0                                                     & \frac{\partial      i}{\partial \left(a/R_s\right)} & \frac{\partial      i}{\partial b} \\
0 & \frac{\partial     d_{tr}}{\partial P} & \frac{\partial d_{tr}}{\partial \left(R_p/R_s\right)} & \frac{\partial d_{tr}}{\partial \left(a/R_s\right)} & \frac{\partial d_{tr}}{\partial b} \\
0 & \frac{\partial     d_{in}}{\partial P} & \frac{\partial d_{in}}{\partial \left(R_p/R_s\right)} & \frac{\partial d_{in}}{\partial \left(a/R_s\right)} & \frac{\partial d_{in}}{\partial b} \\
0 & 0                                      & \frac{\partial      D}{\partial \left(R_p/R_s\right)} & \frac{\partial      D}{\partial \left(a/R_s\right)} & \frac{\partial      D}{\partial b} \\
0 & \frac{\partial     T_{eq}}{\partial P} & 0                                                     & 0                                                   & 0                                  \\
0 & \frac{\partial \phi_{eff}}{\partial P} & 0                                                     & 0                                                   & 0                                  \\
\end{array}
\right] \; \mathrm{and}
\end{equation}

\begin{equation}
\frac{\partial \pmb{\psi}}{\partial \pmb{\alpha}} = 
\left[
\begin{array}{ccc}
\frac{\partial       R_p}{\partial R_s}  & 0                                      & 0                                            \\
\frac{\partial         a}{\partial R_s}  & \frac{\partial          a}{\partial g} & 0                                            \\
0                                        & 0                                      & 0                                            \\
0                                        & 0                                      & 0                                            \\
0                                        & 0                                      & 0                                            \\
0                                        & 0                                      & 0                                            \\
\frac{\partial     T_{eq}}{\partial R_s} & \frac{\partial     T_{eq}}{\partial g} & \frac{\partial     T_{eq}}{\partial T_{eff}} \\
\frac{\partial \phi_{eff}}{\partial R_s} & \frac{\partial \phi_{eff}}{\partial g} & \frac{\partial \phi_{eff}}{\partial T_{eff}} \\
\end{array}
\right]. 
\end{equation}

Note that the derived parameters $i$, $d_{tr}$, $d_{in}$, and $D$ are determined independently of the stellar parameters; therefore, their partial derivatives with respect to the stellar parameters are all identically zero. 

Since the transit depth $D$ is determined from the model light curve generated by the geometric transit signal generator, the elements $\partial D / \partial \left(R_p/R_s\right)$, $\partial D / \partial \left(a/R_s\right)$, and $\partial D / \partial b$ of the Jacobian $\partial \pmb{\psi} / \partial \pmb{\theta}$ are determined numerically. The other non-zero elements of the Jacobians $\partial \pmb{\psi} / \partial \pmb{\theta}$, and $\partial \pmb{\psi} / \partial \pmb{\alpha}$ are calculated according to the following equations:

\begin{equation}
\frac{\partial R_p}{\partial \left(R_p/R_s\right)} = \frac{R_p}{\left(R_p/R_s\right)},
\end{equation}

\begin{equation}
\frac{\partial a}{\partial P} = \frac{2}{3} \, \frac{a}{P},
\end{equation}

\begin{equation}
\frac{\partial i}{\partial \left(a/R_s\right)} = \frac{180}{\pi} \, \frac{b}{\left(a/R_s\right)} \, \frac{1}{\sqrt{\left(a/R_s\right)^2 - b^2}},
\end{equation}

\begin{equation}
\frac{\partial i}{\partial b} = -\frac{180}{\pi} \, \frac{1}{\sqrt{\left(a/R_s\right)^2 - b^2}}, 
\end{equation}

\begin{equation}
\frac{\partial d_{tr}}{\partial P} = \frac{d_{tr}}{P},
\end{equation}

\begin{equation}
\frac{\partial d_{tr}}{\partial \left(R_p/R_s\right)} =  \frac{24 \, P}{\pi} \, 
\frac{1+\left(R_p/R_s\right)}{\sqrt{\left(a/R_s\right)^2 - \left[1+\left(R_p/R_s\right)\right]^2} \, \sqrt{\left[1+\left(R_p/R_s\right)\right]^2 - b^2}},
\end{equation}

\begin{equation}
\frac{\partial d_{tr}}{\partial \left(  a/R_s\right)} = -\frac{24 \, P}{\pi} \, \frac{\left(a/R_s\right)}{ \left(a/R_s\right)^2 - b^2} \,
\frac{\sqrt{\left[1+\left(R_p/R_s\right)\right]^2 - b^2}}{\sqrt{\left(a/R_s\right)^2 - \left[1+\left(R_p/R_s\right)\right]^2}},
\end{equation}

\begin{equation}
\frac{\partial d_{tr}}{\partial b}                    = -\frac{24 \, P}{\pi} \, \frac{b}{ \left(a/R_s\right)^2 - b^2} \,
\frac{\sqrt{\left(a/R_s\right)^2 - \left[1+\left(R_p/R_s\right)\right]^2}}{\sqrt{\left[1+\left(R_p/R_s\right)\right]^2 - b^2}}, 
\end{equation}

\begin{equation}
\frac{\partial d_{in}}{\partial P} = \frac{d_{in}}{P},
\end{equation}

\begin{equation}
\begin{split}
\frac{\partial d_{in}}{\partial \left(R_p/R_s\right)} =  \frac{12 \, P}{\pi} \, 
\left( \frac{1+\left(R_p/R_s\right)}{\sqrt{\left(a/R_s\right)^2 - \left[1+\left(R_p/R_s\right)\right]^2} \, \sqrt{\left[1+\left(R_p/R_s\right)\right]^2 - b^2}} +  \right. \\
\left. \frac{1-\left(R_p/R_s\right)}{\sqrt{\left(a/R_s\right)^2 - \left[1-\left(R_p/R_s\right)\right]^2} \, \sqrt{\left[1-\left(R_p/R_s\right)\right]^2 - b^2}} \right)
\end{split},
\end{equation}

\begin{equation}
\begin{split}
\frac{\partial d_{in}}{\partial \left(  a/R_s\right)} = -\frac{12 \, P}{\pi} \, \frac{\left(a/R_s\right)}{\left(a/R_s\right)^2 - b^2} \,
\left( \frac{\sqrt{\left[1+\left(R_p/R_s\right)\right]^2 - b^2}}{\sqrt{\left(a/R_s\right)^2 - \left[1+\left(R_p/R_s\right)\right]^2}} -  \right.\\
\left. \frac{\sqrt{\left[1-\left(R_p/R_s\right)\right]^2 - b^2}}{\sqrt{\left(a/R_s\right)^2 - \left[1-\left(R_p/R_s\right)\right]^2}} \right)
\end{split},
\end{equation}

\begin{equation}
\begin{split}
\frac{\partial d_{in}}{\partial b} = -\frac{12 \, P}{\pi} \, \frac{b}{\left(a/R_s\right)^2 - b^2} \,
\left( \frac{\sqrt{\left(a/R_s\right)^2 - \left[1+\left(R_p/R_s\right)\right]^2}}{ \sqrt{\left[1+\left(R_p/R_s\right)\right]^2 - b^2}} - \right. \\
\left. \frac{\sqrt{\left(a/R_s\right)^2 - \left[1-\left(R_p/R_s\right)\right]^2}}{ \sqrt{\left[1-\left(R_p/R_s\right)\right]^2 - b^2}} \right)
\end{split},
\end{equation}

\begin{equation}
\frac{\partial T_{eq}}{\partial P} = -\frac{1}{3} \, \frac{T_{eq}}{P},
\end{equation}

\begin{equation}
\frac{\partial \phi_{eff}}{\partial P} = -\frac{4}{3} \, \frac{\phi_{eff}}{P},
\end{equation}

\begin{equation}
\frac{\partial R_p}{R_s} = \frac{R_p}{R_s},
\end{equation}

\begin{equation}
\frac{\partial a}{R_s} = \frac{2}{3} \, \frac{a}{R_s},
\end{equation}

\begin{equation}
\frac{\partial a}{g} = \frac{1}{3} \, \frac{a}{g},
\end{equation}

\begin{equation}
\frac{\partial T_{eq}}{R_s} = \frac{1}{6} \, \frac{T_{eq}}{R_s},
\end{equation}

\begin{equation}
\frac{\partial T_{eq}}{g} = -\frac{1}{6} \, \frac{T_{eq}}{g},
\end{equation}

\begin{equation}
\frac{\partial T_{eq}}{T_{eff}} = \frac{T_{eq}}{T_{eff}},
\end{equation}

\begin{equation}
\frac{\partial \phi_{eff}}{R_s} = \frac{2}{3} \, \frac{\phi_{eff}}{R_s},
\end{equation}

\begin{equation}
\frac{\partial \phi_{eff}}{g} = -\frac{2}{3} \, \frac{\phi_{eff}}{g}, \;\mathrm{and}
\end{equation}

\begin{equation}
\frac{\partial \phi_{eff}}{T_{eff}} = 4 \, \frac{\phi_{eff}}{T_{eff}}.
\end{equation}

% ***********************************************************************************************************

% ***********************************************************************************************************

\end{document}